\documentclass[journal,draftcls,onecolumn,twoside,12pt]{IEEEtran}
\normalsize
\ifCLASSINFOpdf
\else
\fi

\usepackage{amsmath}
\usepackage{cite}
\usepackage{graphicx}
\usepackage{caption}
\usepackage{subcaption}
\usepackage{float}
\usepackage{setspace}
\usepackage{multirow}
\restylefloat{figure}
\input{epsf}
\begin{document}
\doublespacing

\title{On Capacity and Capacity per Unit Cost of Gaussian Multiple Access Channel with Peak Power Constraints}

\author{Siavash~Ghavami,~\IEEEmembership{Student~Member,~IEEE,}
        and~Farshad~Lahouti,~\IEEEmembership{Senior~Member,~IEEE}% <-this % stops a space
\thanks{S. Ghavami and F. Lahouti are with the School
of Elec.~and Comp.~Eng., University of Tehran, Tehran,
Iran, e-mail: (s.ghavami,lahouti)@ut.ac.ir.}% <-this % stops a space
% <-this % stops a space
%\thanks{Mr.~Ghavami was supported in part by the Iranian Ministry of Science, Research and Technology  and in part by the National Science and Engineering Research Council (NSERC) Canada.}
}

\markboth{}%
{}

\maketitle
\vspace*{-1.in}
\begin{abstract}
%\boldmath
This paper investigates the capacity and capacity per unit cost of Gaussian multiple-access channel (GMAC) with peak power constraints. We first devise an approach based on Blahut-Arimoto Algorithm to numerically optimize the sum rate and quantify the corresponding input distributions. The results reveal that in the case with identical peak power constraints, the user with higher SNR is to have a symmetric antipodal input distribution for all values of noise variance. Next, we analytically derive and characterize an achievable rate region for the capacity in cases with small peak power constraints, which coincides with the capacity in a certain scenario. The capacity per unit cost is of interest in low power regimes and is a target performance measure in energy efficient communications. In this work, we derive the capacity per unit cost of additive white Gaussian channel and GMAC with peak power constraints. The results in case of GMAC demonstrate that the capacity per unit cost is obtained using antipodal signaling for both users and is independent of users’ rate ratio. We characterize the optimized transmission strategies obtained for capacity and capacity per unit cost with peak-power constraint in detail and specifically in contrast to the settings with average-power constraints.
\end{abstract}

\vspace*{-0.25in}

\IEEEpeerreviewmaketitle

\section{Introduction}
There is growing attention recently towards energy efficient transmission in modern so-called green communication systems. From an information theoretic view point, green power communications can be interpreted in different ways. In~\cite{1,2,3}, to devise optimized reliable transmission strategies, in addition to the transmission power, the expended energy for processing the data at the encoder and the decoder is also taken into account. Other perspectives in green power communications include peak power constrained transmission~\cite{4}, and considering capacity per unit cost~\cite{5} for reduced energy expenditure of data transmission. It is conjectured in~\cite{6,7}, that neurons set their transmission strategy to achieve the capacity per unit cost~\cite{5}, and hence facilitate Petabit per second communications in an energy efficient manner. In this paper, we investigate the energy efficient strategy for transmission over MAC with peak power constraints.

In the literature, certain problems in transmission strategy design for achieving the capacity per unit cost and the capacity of certain channels with peak power constraint have been previously studied. In the following two subsections, the related works are briefly reviewed.

\subsection{Related works in capacity per unit cost}
A trade off exists between information rate in terms of bit per second (bps) and information per unit cost in terms of bit per joule (bpj) in the band limited channel with average transmitted power constraint and additive white Gaussian noise (AWGN). The celebrated formula for capacity of AWGN channel~\cite{8} is a logarithmic function in terms of signal to noise ratio (SNR). It is evident that by increasing the power, the slope of the capacity function (in terms of bpj) is reduced. In other words, as data rate increases, it is not avoidable that the energy efficiency of communication is reduced. This result is not only limited to the AWGN channel, but it applies to any channel with a power constraint~\cite{9}.

In communication systems with large available bandwidth and limited energy budget, minimizing the energy per bit or equivalently achieving the capacity per unit cost could be more critical in comparison with maximizing the communication rates. The minimum energy per bit of point-to-point channels with additive noise is studied in wide generality in~\cite{10}. The minimum energy per bit is also known for the Gaussian multiple-access channel (GMAC), the Gaussian broadcast channel and the Gaussian interference channel~\cite{5,10,11,12}. However, the minimum energy per bit is not yet known for the three terminal setting of a Gaussian relay channel, though some progress has been made (the interested reader is referred to~\cite{13,14} and the references therein). More results on the capacity per unit cost of the three terminal channels, relay networks and arbitrary networks are reported in~\cite{15,16} and~\cite{17}. Recently in~\cite{18}, the wideband slope of outage capacity is quantified in several Gaussian communications settings. In~\cite{5}, it is shown that for  MAC with infinite bandwidth, the capacity per unit cost is achieved using time division multiple access (TDMA). In~\cite{12}, for MAC with finite (but large) bandwidth, the optimal transmission strategy is shown to deviate from TDMA.

\subsection{Related works in capacity with peak power constraint}
As stated, another perspective to green power communications is the transmission with limited peak power. This is also of great practical importance as it describes situations where the transmitter has a limited output back off~\cite{19}. Peak power constraint at mobile users is one of practical limitation in uplink scenario of mobile communication~\cite{20,21}. A general solution for the capacity of point to point AWGN channel with peak power constraint is not available. However, it is shown that for a given value of peak power constraint, the capacity-achieving distribution is unique and discrete with finite symmetric support region with center on zero~\cite{22}. In addition, for peak power values less than 1.05 (with unit noise variance), the capacity is achieved by equiprobable antipodal signaling~\cite{23}. This is proven based on the relation of mutual information in Gaussian channels to minimum mean-squared error estimation~\cite{24}. The low SNR capacity of point to point fading channel with peak and average power constraints is considered in~\cite{25}. In~\cite{26} by considering optical communications, an analytical closed form expression for a capacity-approaching input distribution is developed via input entropy maximization under non-negativity, peak and average power constraints. The low-SNR capacity of single-input single-output and multiple-input multiple-output non-coherent fading channels with peak and average power constraints is studied in~\cite{27}. In~\cite{28} the capacity-achieving probability measure under boundedness constraint and average cost constraints for conditionally Gaussian channels is studied. In~\cite{29}, bounds are derived on the non-coherent capacity of wide-sense stationary uncorrelated scattering channels that are selective both in time and frequency and are under-spread. The aim is to quantify the capacity-optimal bandwidth as a function of the peak power and the channel scattering function. In neuronal communications, a noisy spiking neuron has been shown to maximize mutual information within given range of firing rate or peak power constraint ~\cite{28,30,31}.

\subsection{Outline of Contributions}
In this paper, we investigate the capacity and the capacity per unit cost of Gaussian-multiple access channel with peak power constraints. Prior work on the capacity of GMAC with peak power constraints is reported in~\cite{32,33,34}, where it is shown that generating the codebooks of users according to discrete distributions achieves the largest sum rate of two-user GMAC. While~\cite{34} proves the discreteness for the sum rate,~\cite{35} proves the discreteness for the entire rate region. Autuors in~\cite{35} also gives results for the stochastic amplitude constraint case. The capacity per unit cost with peak power constraint is only considered in~\cite{36} for point to point fading channels with or without channel state information. Our search did not reveal any prior art on capacity per unit cost of GMAC with peak power constraints.
In Section II, we present an approach based on the Blahut-Arimoto (BA) algorithm to numerically optimize an achievable sum rate of GMAC with peak power constraints. This is accomplished by identifying the corresponding users codebooks input distributions. Our results reveal that for the case with identical peak power constraints, the input distribution of the user with higher signal to noise ratio (SNR) is equiprobable antipodal for all values of noise variance. In the limiting regimes of low and high transmission powers, the optimized distributions of users quantify to antipodal distributions for both users and an antipodal distribution and a uniform distribution, respectively. The latter is consistent with the results of~\cite{37}.

In Section III, we derive the capacity per unit cost of additive white Gaussian channel and GMAC with peak power constraints. The results in case of GMAC demonstrate that the capacity per unit cost is obtained using antipodal signaling and a successive decoding strategy. Interestingly, the associated slope rate region turns out to be independent of users’ rate ratio. We characterize the optimized transmission strategies obtained for capacity and capacity per unit cost with peak power constraints in detail and specifically in contrast to the settings with average power constraints. This is tabulated in concluding remarks of Section IV. The study of the capacity per unit cost of communication networks with peak power constraints could shed light on efficient transmission strategies for next generation green communication systems.

\section{Achievable Sum Rate of GMAC with Peak Power Constraints}
In this Section, we first derive the input distributions of users 1 and 2 for maximizing sum rate of GMAC with peak power constraints numerically in different SNR regimes. Second, we derive an achievable rate region for GMAC in low SNR regime with antipodal distribution for inputs, which maximizes the sum rate numerically. This achievable region used in the next Section of this paper in deriving capacity per unit cost. Last in this Section, the capacity region of GMAC with infinite bandwidth obtained.

The output of GMAC with peak power constraints is described by
\begin{equation}\label{eq:1}
Y = {X_1} + {X_2} + Z,
\end{equation}
where ${X_i}$, $i \in \left\{ {1,2} \right\}$ are real-valued random inputs, ${X_i} \in \left[ { - \sqrt {{\rho _i}} ,\sqrt {{\rho _i}} } \right]$ , and $Z\sim{{\cal N}}\left( {0,{\sigma ^2} = 1} \right)$ is the independent additive white Gaussian noise.
 \vspace{-0.5\baselineskip}
\subsection{Optimum Input Distribution for Maximization of Sum Rate}
We consider maximizing the sum rate of GMAC with peak power constraints, where the optimal input distributions are known to be discrete~\cite{32}. The desired optimization problem with respect to the input distributions is expressed as follows,
\begin{subequations}\label{eq:2}
\begin{align}
\label{eq:2a}
&\mathop {\sup }\limits_{p({x_i}):\left| {{X_i}} \right| \le \sqrt {{\rho _i}} ,i = 1,2} I\left( {{X_1},{X_2};Y} \right) = \\ \label{eq:2b}
&\mathop {\sup }\limits_{p({x_i}),:\left| {{X_i}} \right| \le \sqrt {{\rho _i}} ,i = 1,2} h\left( {{X_1} + {X_2} + Z} \right)
\end{align}
\end{subequations}
where $p\left( {{x_i}} \right) = p\left( {{X_i} = {x_i}} \right)$ is discrete input probability distributions, and given by
\begin{equation}\label{eq:3}
p\left( {{x_i}} \right) = \sum\limits_{j = 1}^{{K_1}} {{a_{ij}}\delta \left( {{x_i} - {{b}_{ij}}} \right)} {\rm{,}}\,{K_i} \in {Z^ + }, i\in\{1,2\},
\end{equation}
where ${a_{ij}} = p({x_i} = {b_{ij}})$, and ${b_{ij}}$ denotes $j$th mass point for ${X_i}$. In~\eqref{eq:3}, ${K_i}$ is the number of probability mass points for user $i$, and ${Z^ + }$ is the set of positive integers. Using some simple manipulations and due to the independency of ${X_1}$  and ${X_2}$, \eqref{eq:2b} may be rewritten as follows
\begin{subequations}\label{eq:4}
\begin{align}
\label{eq:4a}
&\mathop {\sup }\limits_{{{b}_{1j}},{{b}_{2j}},{a_{1j}},{a_{2k}},{K_1},{K_2}} \frac{{ - 1}}{{\sqrt {2\pi {\sigma ^2}} }}\int_{ - \infty }^\infty  {\sum\limits_{j = 1}^{{K_1}} {\sum\limits_{k = 1}^{{K_2}} {{a_{1j}}{a_{2k}}\exp \left( { - \frac{{{{\left( {y - {{b}_{1j}} - {{b}_{2k}}} \right)}^2}}}{{2{\sigma ^2}}}} \right)} } } \times \\ \nonumber & \log  \left( {\frac{1}{{\sqrt {2\pi {\sigma ^2}} }}\sum\limits_{j = 1}^{{K_1}} {\sum\limits_{k = 1}^{{K_2}} {{a_{1j}}{a_{2k}}\exp \left( { - \frac{{{{\left( {y - {{b}_{1j}} - {{b}_{2k}}} \right)}^2}}}{{2{\sigma ^2}}}} \right)} } } \right)dy,\\ \nonumber s.t.
\\ \label{eq:4b}
& \left| {{{b}_{1j}}} \right| \le \sqrt {{\rho _1}} ,0 \le {a_{1j}} \le 1,{\rm{  }}j \in \left[ {1,{K_1}} \right],
\\ \label{eq:4c}
& \left| {{{b}_{2k}}} \right| \le \sqrt {{\rho _2}} ,0 \le {a_{2k}} \le 1,{\rm{ }}k \in \left[ {1,{K_2}} \right].
\end{align}
\end{subequations}
Simply, by double differentiating with respect to ${a_{1j}}$ and ${a_{2k}}$ it can be shown that the objective function in~\eqref{eq:4a} is concave with respect to ${a_{1j}}$  and ${a_{2k}}$. A closed form solution for the above optimization problem may not be obtained in general. We select, ${K_i}$ enough large to ${b_{1j}}$  and ${b_{2k}}$ cover all region of $\left[ { - \sqrt {{\rho _i}} ,\sqrt {{\rho _i}} } \right]$  with good approximation, Hence optimization problem is simplified to finding $a_{i,j}$, $i,j \in \{1,2\}$.

Here, we present the alternating maximization procedure on which our algorithm is based. Let
\begin{eqnarray}\label{eq:5}
f\left( {{{\bf{a}}_1},{{\bf{a}}_2}} \right) =  - \frac{1}{{\sqrt {2\pi {\sigma ^2}} }}\int_{ - \infty }^\infty  {\sum\limits_{j = 1}^{{K_1}} {\sum\limits_{k = 1}^{{K_2}} {{a_{1j}}{a_{2k}}\exp \left( { - \frac{{{{\left( {y - {{b}_{1j}} - {{b}_{2k}}} \right)}^2}}}{{2{\sigma ^2}}}} \right)} } } \times \\ \nonumber \log  \left( {\frac{1}{{\sqrt {2\pi {\sigma ^2}} }}\sum\limits_{j = 1}^{{K_1}} {\sum\limits_{k = 1}^{{K_2}} {{a_{1j}}{a_{2k}}\exp \left( { - \frac{{{{\left( {y - {{b}_{1j}} - {{b}_{2k}}} \right)}^2}}}{{2{\sigma ^2}}}} \right)} } } \right)dy
\end{eqnarray}
where ${{\bf{a}}_i} = \left[ {{a_{ij}}} \right]$, $1 \le j \le {K_i}$, and let us consider the optimization problem given by
\begin{equation}
\mathop {\sup }\limits_{{{\bf{a}}_1} \in {A_1},{{\bf{a}}_2} \in {A_2}} f\left( {{{\bf{a}}_1},{{\bf{a}}_2}} \right) = {f^*}
\end{equation}
where ${A_1}$  and ${A_2}$  are the sets we optimize over. We denote by ${c_2}\left( {{{\bf{a}}_1}} \right)$  the point that achieves $\mathop {\sup }\limits_{{{\bf{a}}_2} \in {A_2}} f\left( {{{\bf{a}}_1},{{\bf{a}}_2}} \right)$ , and by ${c_1}\left( {{{\bf{a}}_2}} \right)$  the one that achieves $\mathop {\sup }\limits_{{{\bf{a}}_1} \in {A_1}} f\left( {{{\bf{a}}_1},{{\bf{a}}_2}} \right)$. The algorithm is performed by iterations, where in each iteration we maximize over one of the variables. Let $\left( {{\bf{a}}_1^{(0)},{\bf{a}}_2^{(0)}} \right)$ be an arbitrary point in ${A_1} \times {A_2}$. For $r \ge 0$ let
\[\left( {{\bf{a}}_1^{(r)},{\bf{a}}_2^{{(r)}}} \right) = \left( {{c_1}\left( {{\bf{a}}_2^{\left( {r - 1} \right)}} \right),{c_2}\left( {{c_1}\left( {{\bf{a}}_2^{\left( {r - 1} \right)}} \right)} \right)} \right)\]
and let ${f^{(r)}} = f\left( {{\bf{a}}_1^{(r)},{\bf{a}}_2^{(r)}} \right)$ be the value in the current iteration. The following lemma describes the conditions the problem needs to meet to ${f^{(r)}}$ converges to ${f^*}$  by approaching $r$ to infinity.

\emph{Lemma 1}. \label{lemma:1} Convergence of the alternating maximization procedure(~\cite{38} ,Lemmas 9.4 and 9.5): Let $f\left( {{{\bf{a}}_1},{{\bf{a}}_2}} \right)$  be a real, concave, bounded-from-above function that is continuous and has continuous partial derivatives, and let the sets ${A_1}$ and ${A_2}$ which we maximize over those,  be convex, which is due to linear constraints on (4-b) and (4-c). Further, assume that ${c_2({\bf{a}}_1}) \in {A_2}$ and $c_1({{\bf{a}}_2}) \in {A_1}$ for all ${{\bf{a}}_1} \in {A_1}$ , ${{\bf{a}}_2} \in {A_2}$ and that ${c_2}\left( {{{\bf{a}}_1}} \right)$, ${c_1}\left( {{{\bf{a}}_2}} \right)$  are unique. Under these conditions, $\mathop {\lim }\limits_{r \to \infty } {f^{(r)}} = {f^*}$, where ${f^*}$ is the global maximum.

\begin{figure}[t!]
        \centering
        \begin{subfigure}[b]{0.49\textwidth}
                \caption{}
                \includegraphics[width=\textwidth]{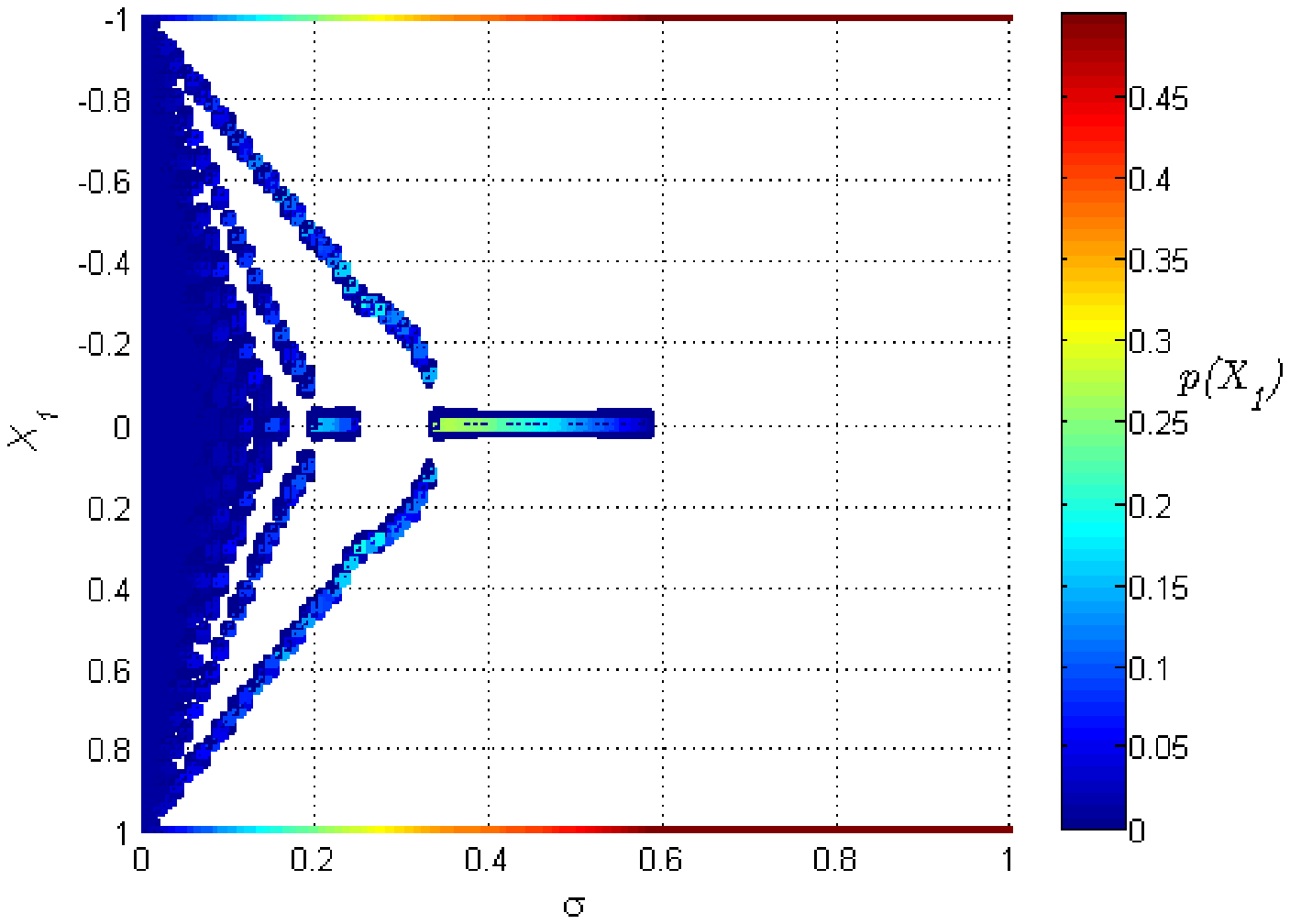}
                \label{fig:p_user1}
        \end{subfigure}%
        ~ %add desired spacing between images, e. g. ~, \quad, \qquad, \hfill etc.
        \begin{subfigure}[b]{0.49\textwidth}
                \caption{}
                \includegraphics[width=\textwidth]{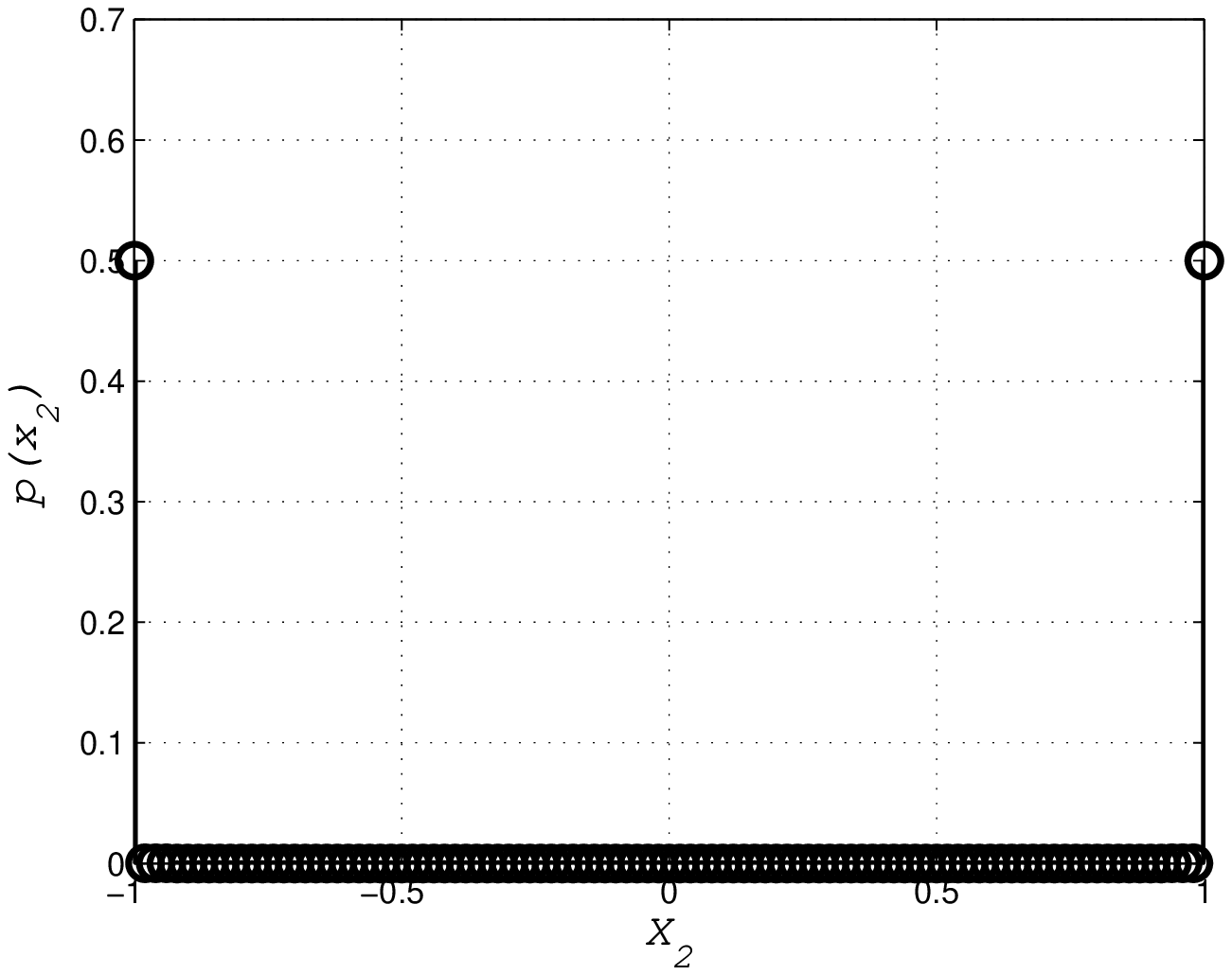}
                \label{fig:p_user2}
        \end{subfigure}
        \\
        \vspace{-1.5\baselineskip}
        \caption{(a) $p({x_1})$ maximizing ${R_1} + {R_2}$ in terms of ${X_1}$ and $\sigma $; (b) $p({x_2})$ maximizing ${R_1} + {R_2}$ for all values of $\sigma $.}
        \label{fig:ave_con}
\end{figure}

Below, we present an iterative algorithm developed based on the above lemma using the BA algorithm~\cite{9,40} and generalized BA algorithm for discrete memoryless MAC (DMC-MAC)~\cite{41}, and explore the sum rate of GMAC with peak power constraints numerically. Specifically, the proposed algorithm aims at an optimized achievable sum rate based on a successive decoding strategy. The details of the proposed algorithm are as follows%and considering user 2 as the one having a higher SNR that is decoded first.
\begin{enumerate}
  \item We consider $K_i$ equidistant mass points over the support region of ${X_i}$ , $i \in \left\{ {1,2} \right\}$ , ($K_i$ very large). The probabilities of all mass points for  ${X_i}$s are initialized randomly.
  \item Considering the sum rate ${R_1} + {R_2} \le I\left( {{X_1},{X_2};Y} \right)$, we aim to maximize $I\left( {Y;{X_1},{X_2}} \right)$  with respect to $p\left( {{x_1}} \right)$, when $p\left( {{x_2}} \right)$ is given. The sequence of ${p^{(r)}}\left( {{x_1}} \right)$, $r = 0,1,...$ is given by~\cite{41}
      \begin{equation}\label{eq:7}
      p_{}^{(r + 1)}\left( {{x_1}} \right) = p_{}^{(r)}\left( {{x_1}} \right)\frac{{\exp \left( {{I_1}\left( {{x_1};Y} \right)} \right)}}{{\sum\nolimits_{{{x'}_1}} {p_{}^{(r)}\left( {{{x'}_1}} \right)\exp \left( {{I_1}\left( {{{x'}_1};Y} \right)} \right)} }}
      \end{equation}
      where $p_{}^{(0)}\left( {{x_1}} \right) \ne 0$  and
      \begin{equation}\label{eq:8}
      {I_1}\left( {{x_1};Y} \right) = \sum\nolimits_{{x_2}} {{p^{\left( r \right)}}\left( {{x_2}} \right)I\left( {{x_1},{x_2};Y} \right)}
      \end{equation}
      where $I\left( {{x_1},{x_2};Y} \right)$ is described by
      \begin{equation}\label{eq:9}
      I\left( {{x_1},{x_2};Y} \right) = D\left( {p\left( {\left. Y \right|{x_1},{x_2}} \right)||{p^{\left( r \right)}}\left( Y \right)} \right).
      \end{equation}
      In~\eqref{eq:9} $D\left( {.|.} \right)$ is Kullback-Leibler distance and we have
      \begin{equation}\label{eq:10}
      D\left( {p\left( {\left. Y \right|{x_1},{x_2}} \right)||p^{(r)}\left( Y \right)} \right) = \int\limits_{ - \infty }^\infty  {p\left( {\left. y \right|{x_1},{x_2}} \right)\ln \frac{{p\left( {\left. y \right|{x_1},{x_2}} \right)}}{{p^{(r)}\left( y \right)}}} dy.
      \end{equation}
      In~\eqref{eq:10}, $p^{(r)}\left( y \right)$ is obtained as
      \begin{equation}\label{eq:11}
      p^{(r)}\left( y \right) = \sum\nolimits_{{x_1}} {\sum\nolimits_{{x_2}} {p_{}^{\left( {r} \right)}\left( {{x_1}} \right)p_{}^{\left( r \right)}\left( {{x_2}} \right)p\left( {\left. y \right|{x_1},{x_2}} \right)}}
      \end{equation}
      and $p^{(r)}\left( {\left. y \right|{x_2}} \right)$, which is used in the next step, obtained by
      \begin{equation}\label{eq:12}
      p^{(r)}\left( {\left. y \right|{x_2}} \right) = \sum\nolimits_{{x_2}} {p^{\left( {r + 1} \right)}\left( {{x_1}} \right)p\left( {\left. y \right|{x_1},{x_2}} \right)}
      \end{equation}
  \item Considering ${R_1} + {R_2} \le I\left( {{X_1},{X_2};Y} \right) = I\left( {{X_2};Y} \right) + I\left( {{X_1};\left. Y \right|{X_2}} \right)$ (based on successive decoding strategy) and a given $p\left( {{x_1}} \right)$, the second term is not a function of $p\left( {{x_2}} \right)$. Hence, we aim to maximize $I\left( {{X_2};Y} \right)$  with respect to $p\left( {{x_2}} \right)$  for a given $p\left( {{x_1}} \right)$. we set up an iterative procedure. Accordingly, $p_{}^{\left( {r + 1} \right)}\left( {{x_2}} \right)$ is given as follows,
      \begin{equation}\label{eq:13}
      p_{}^{(r + 1)}\left( {{x_2}} \right) = p_{}^{(r)}\left( {{x_2}} \right)\frac{{\exp \left( {I\left( {{x_2};Y} \right)} \right)}}{{\sum\nolimits_{{{x'}_2}} {p_{}^r\left( {{{x'}_2}} \right)\exp \left( {I\left( {{{x'}_2};Y} \right)} \right)} }}
      \end{equation}
      where $p_{}^{(0)}\left( {{x_2}} \right) \ne 0$  and $I\left( {{x_2};Y} \right)$ is given by
      \begin{equation}\label{eq:14}
      I\left( {{x_2};Y} \right) = D\left( {\left. {p^{(r)}\left( {\left. Y \right|{x_2}} \right)} \right\|p^{(r)}\left( Y \right)} \right)
      \end{equation}
      where
      \begin{equation}\label{eq:15}
      D\left( {p^{(r)}\left( {\left. Y \right|{x_2}} \right)||p^{(r)}\left( Y \right)} \right) = \int\limits_{ - \infty }^\infty  {p^{(r)}\left( {\left. y \right|{x_2}} \right)\ln \frac{{p^{(r)}\left( {\left. y \right|{x_2}} \right)}}{{p^{(r)}\left( y \right)}}} dy.
      \end{equation}
      %Then $p\left( y \right)$ is obtained as
      \begin{equation}\label{eq:16}
      p^{(r)}\left( y \right) = \sum\nolimits_{{x_1}} {\sum\nolimits_{{x_2}} {p_{}^{\left( {r + 1} \right)}\left( {{x_1}} \right)p_{}^{\left( {r } \right)}\left( {{x_2}} \right)p\left( {\left. y \right|{x_1},{x_2}} \right)}}
      \end{equation}
  \item $I\left( {{X_1},{X_2};Y} \right)$  calculated by $p_{}^{(r)}\left( {{x_1}} \right)$  and $p_{}^{(r)}\left( {{x_2}} \right)$  as distribution of ${X_1}$  and ${X_2}$, are denoted by ${I^{\left( r \right)}}\left( {{X_1},{X_2};Y} \right)$. By this notation, the stopping criterion in the $r^\textit{th}$ iteration is given by
      \begin{equation}\label{eq:17}
      \left| {{I^{\left( {r + 1} \right)}}\left( {{X_1},{X_2};Y} \right) - {I^{\left( r \right)}}\left( {{X_1},{X_2};Y} \right)} \right| < \varepsilon,
      \end{equation}
      where $\varepsilon>0$ is a sufficiently small number. If the above criterion is satisfied, then iteration stops and one concludes with the estimate $p\left( {{x_1}} \right) = p_{}^{(r)}\left( {{x_1}} \right)$ and $p\left( {{x_2}} \right) = p_{}^{(r)}\left( {{x_2}} \right)$ else go to step 2.
\end{enumerate}
Fig~\ref{fig:p_user1} shows $p({x_1})$  in terms of ${x_1}$  and $\sigma $ obtained from the proposed algorithm for maximizing sum rate. It is observed that as the noise standard deviation $\sigma $ decreases, the number of mass points in $p({x_1})$ increases. Mass points around zero appear for $\sigma < 0.34$. Fig~\ref{fig:p_user2} shows $p({x_2})$ maximizing sum rate in terms of ${x_2}$ when $\sqrt {{\rho _1}}  = \sqrt {{\rho _2}}  = 1$. Our numerical results show very importantly that for all values of $\sigma $, $p({x_2})$ has antipodal distribution. For very small values of $\sigma $ ($\sigma  \le 0.003$ almost noiseless), $p({x_1})$  has uniform distribution. This result is consistent with that of~\cite{37} for maximizing the entropy of summation of independent random variables with identical symmetric support regions. Note that, an antipodal distribution for $p({x_2})$ indicate a higher SNR for user 2 when compared to that of user 1 with any $p({x_1})$ and the same peak power constraints. %This is in line with the successive decoding strategy adopted in the proposed algorithm.

\begin{figure}
        \centering
        \begin{subfigure}[b]{0.49\textwidth}
        \caption{}
        \includegraphics[width=\textwidth]{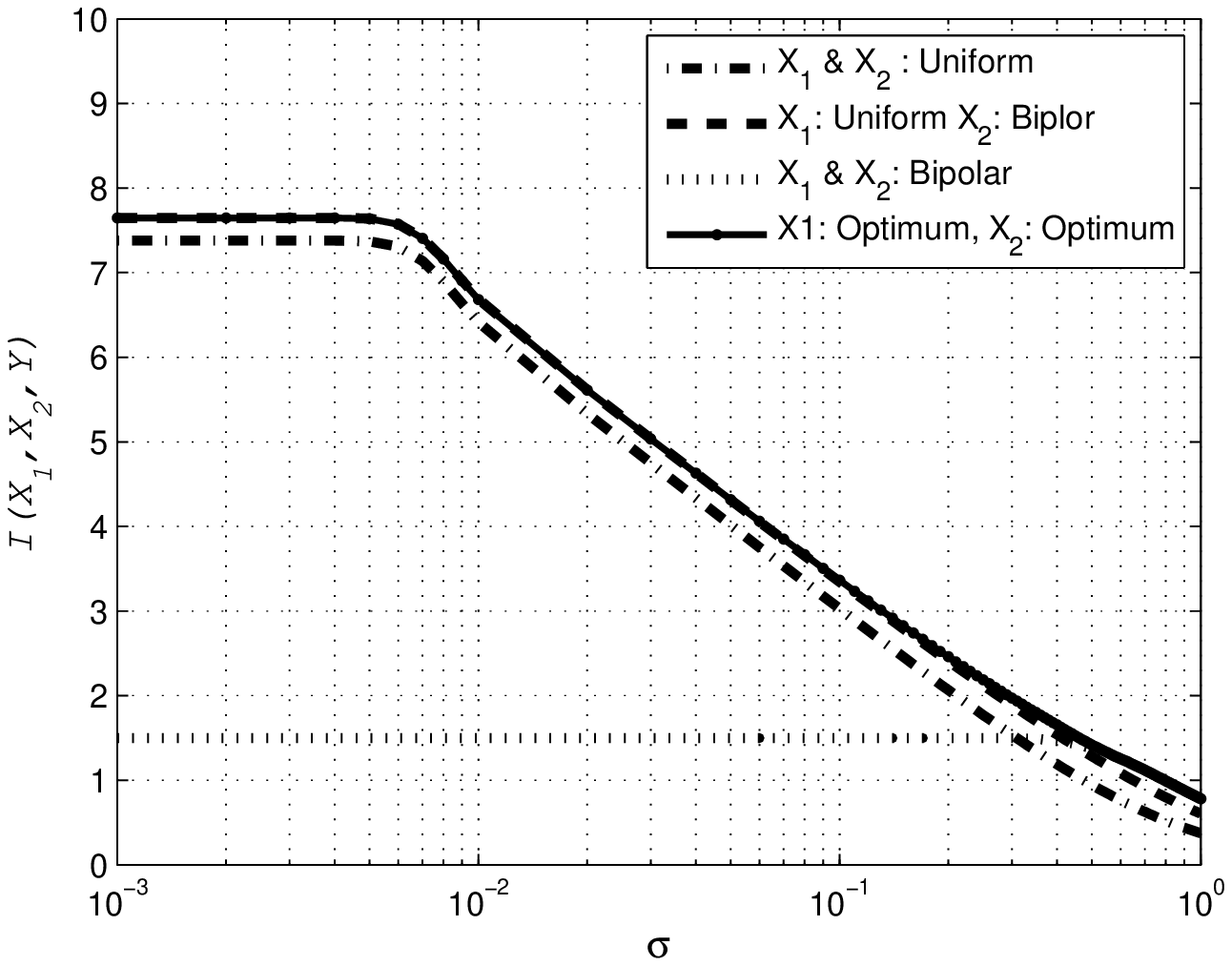}
        \end{subfigure}
        \begin{subfigure}[b]{0.49\textwidth}
        \caption{}
        \includegraphics[width=\textwidth]{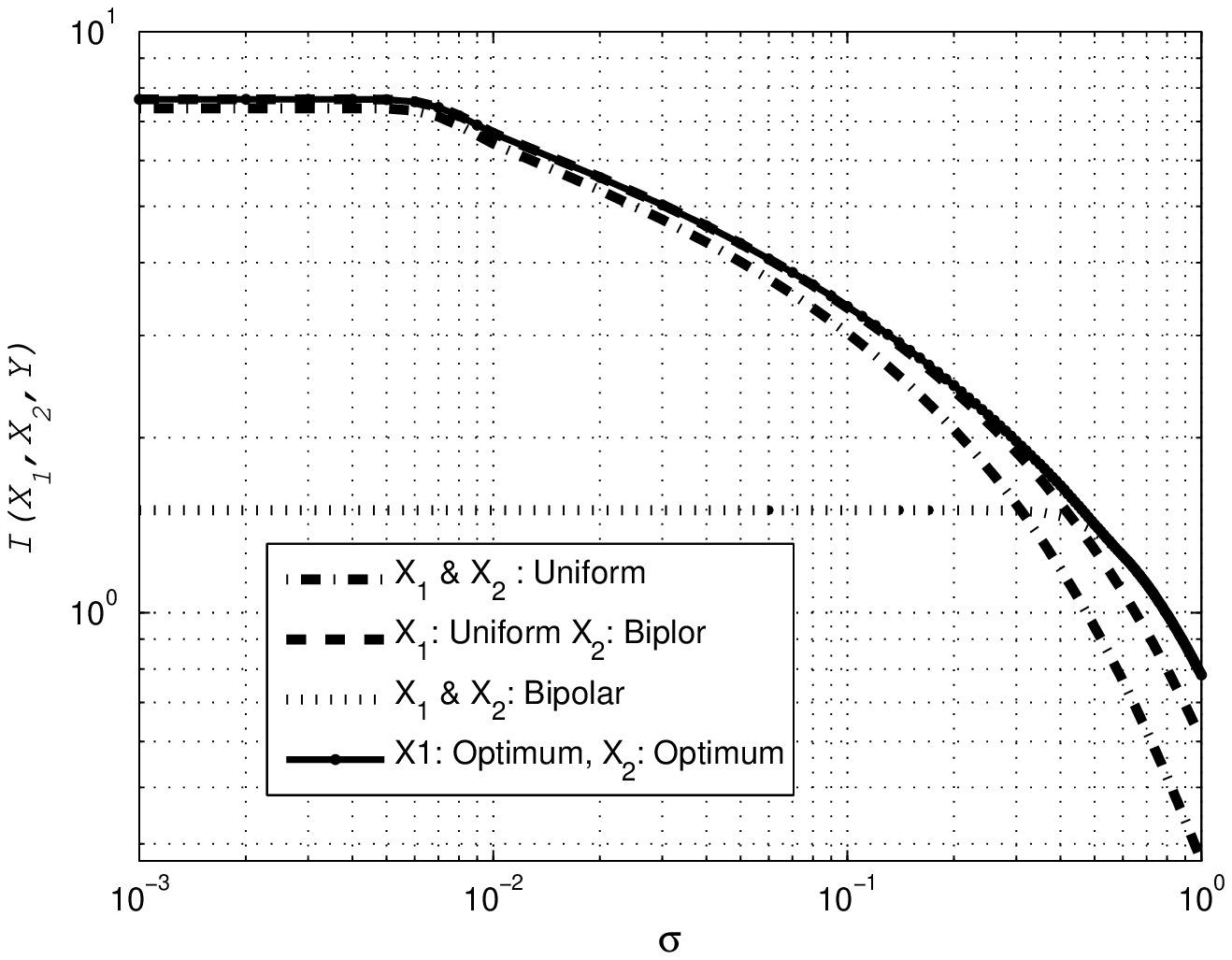}
        \end{subfigure}
        \vspace{-0.5\baselineskip}
        \caption{$I\left( {{X_1},{X_2};Y} \right)$ in terms of $\sigma$ for different input distributions of $X_1$ and $X_2$, (a) linear scale, (b) logarithmic scale.}
        \label{fig:mutual_inf}
\end{figure}

Fig~\ref{fig:mutual_inf} shows $I\left( {{X_1},{X_2};Y} \right)$  in terms of $\sigma $ for different distribution of ${X_1}$ and ${X_2}$. It is demonstrated that the for all values of $\sigma $ the distribution of ${X_1}$ and ${X_2}$ derived from the iterative algorithm, have the largest $I\left( {{X_1},{X_2};Y} \right)$, in comparison with bipolar and/or uniform input distributions. %it converges to $I\left( {{X_1},{X_2};Y} \right)$ with bipolar distributions of ${X_1}$ and ${X_2}$ for large values of $\sigma $, and it converges to $I\left( {{X_1},{X_2};Y} \right)$ with uniform distribution of ${X_1}$  and bipolar distribution of ${X_2}$ for small values of $\sigma $.

\emph{Remark }1. \label{remark:1} The obtained input distribution for user 1, $p({x_1})$, may be interpreted for different values of $\sigma $ as follows. For large values of $\sigma $, an antipodal distribution for $p({x_1})$  maximizes $I\left( {{X_1},{X_2};Y} \right)$ (see Fig~\ref{fig:mutual_inf}). Since this corresponds to a decoding strategy where user 1 treated as noise in step 2, an antipodal distribution may be seen as the Euclidean distance maximizing input in the low SNR regime for smallest errors. As $\sigma $ reduces (SNR increases) a new mass point appears at zero, and the number of alphabets of ${X_1}$ increases for enhanced transmission rate. This distribution is however still the Euclidean distance maximizing input with three mass points and limited peak power. For smaller values of $\sigma $$\left( {\sigma  < 0.1} \right)$, the effect of noise becomes negligible and the input distribution moves towards a uniform distribution for maximized entropy and transmitted information (see Fig~\ref{fig:mutual_inf}).

\emph{Remark} 2. The distribution obtained here for $p({x_1})$ is symmetric around zero and imposes a symmetric Gaussian mixture distributed interference on ${X_2}$  that is treated as noise. This in addition to the Gaussian noise of the channel creates a symmetric additive noise channel for ${X_2}$. Hence, the rate maximizing two mass point distribution of $p({x_2})$ has uniform distribution~\cite{42}.

\begin{figure}[t!]
        \centering
        \begin{subfigure}[b]{0.49\textwidth}
                \caption{}
                \includegraphics[width=\textwidth]{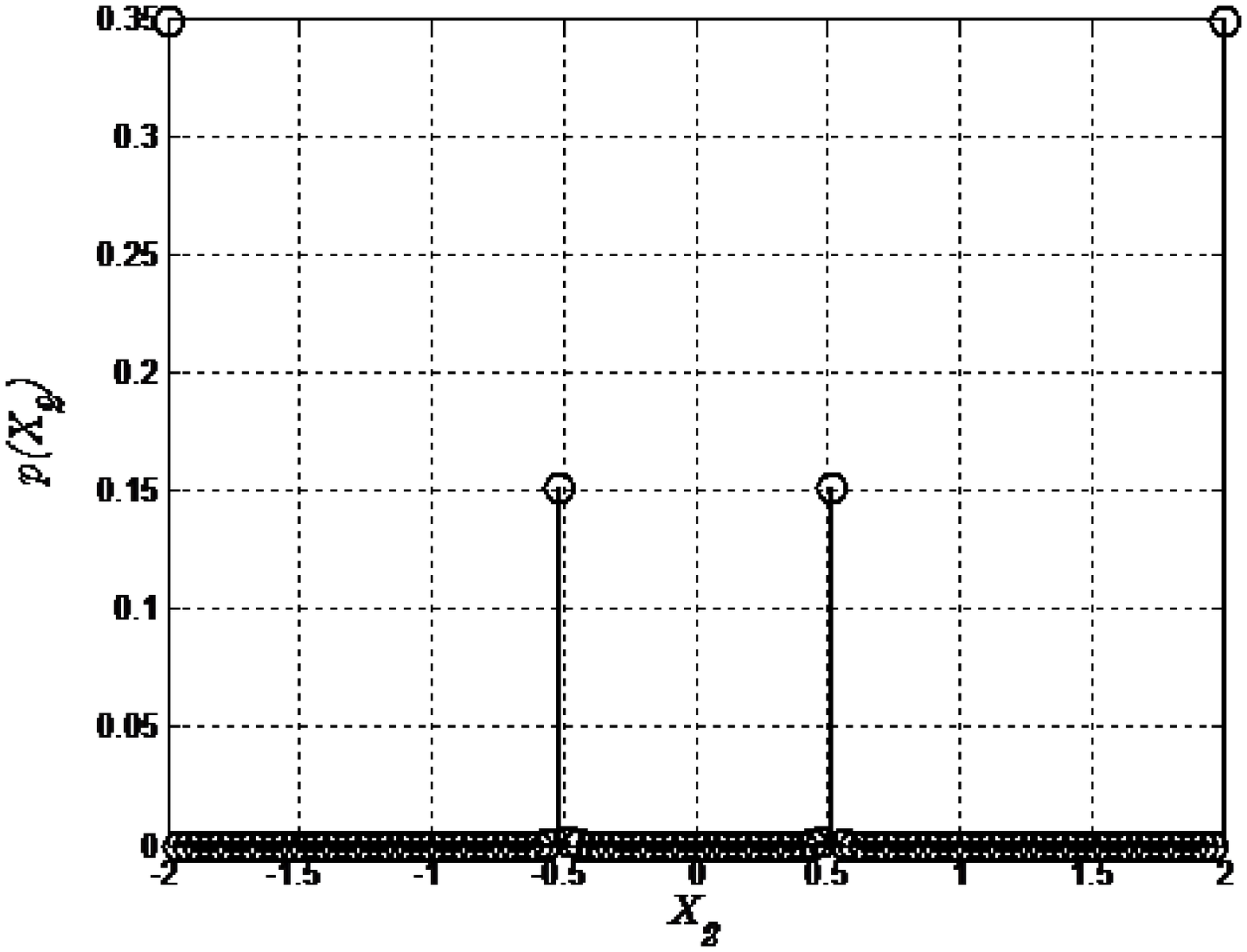}
                \label{fig:Fig3a}
        \end{subfigure}%
        ~ %add desired spacing between images, e. g. ~, \quad, \qquad, \hfill etc.
        \begin{subfigure}[b]{0.49\textwidth}
                \caption{}
                \includegraphics[width=\textwidth]{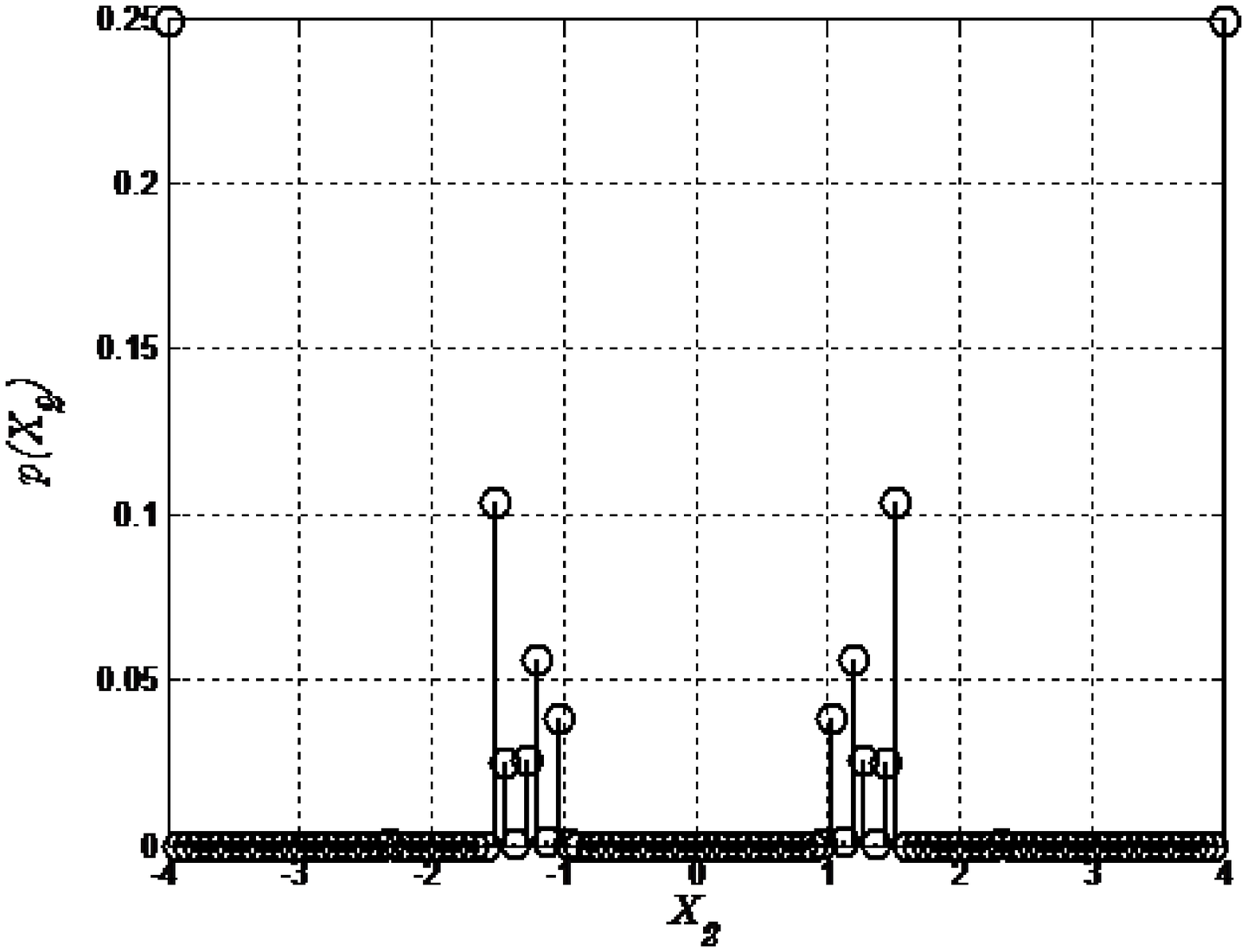}
                \label{fig:Fig3b}
        \end{subfigure}
        \begin{subfigure}[b]{0.49\textwidth}
                \caption{}
                \includegraphics[width=\textwidth]{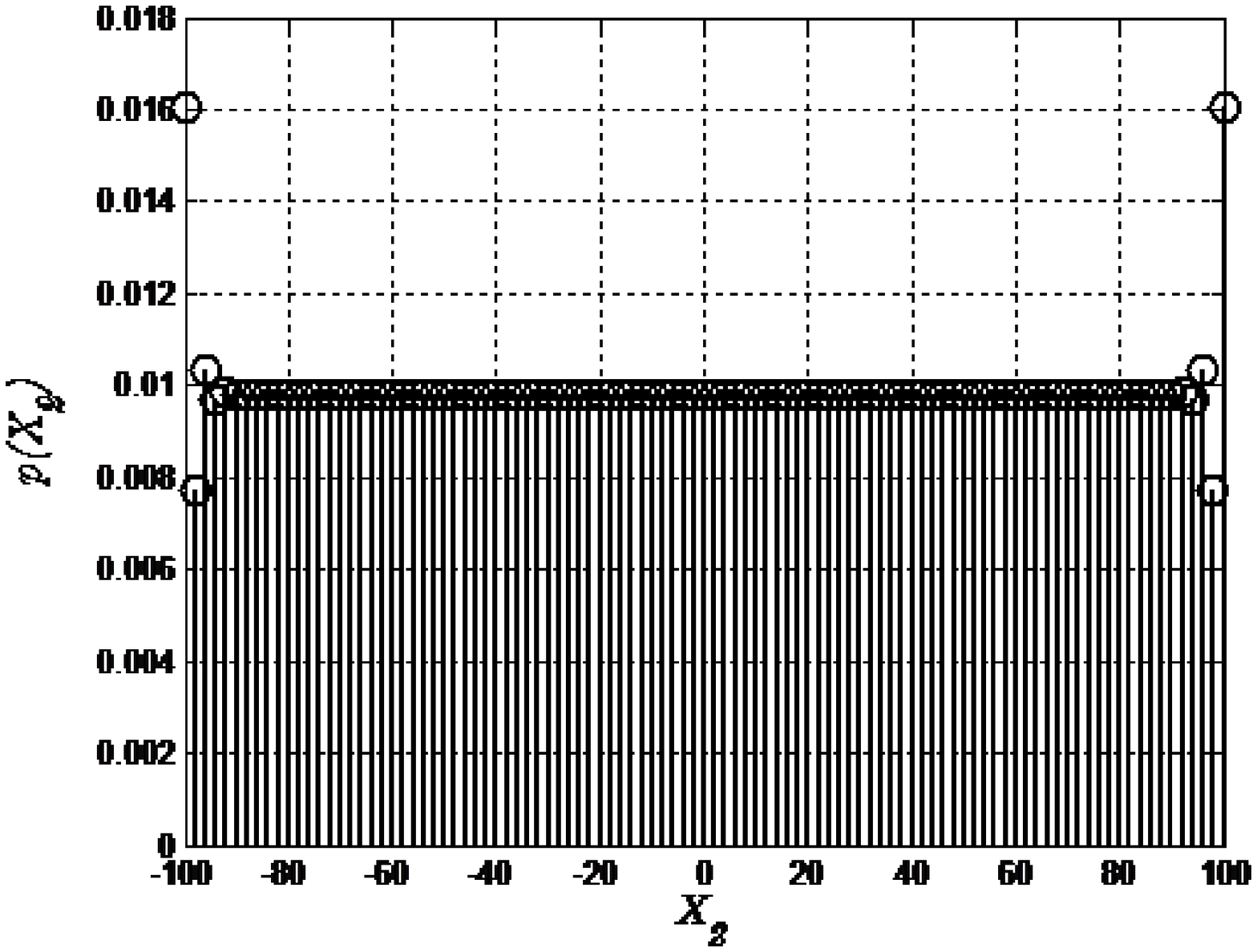}
                \label{fig:Fig3c}
        \end{subfigure}
        \\
        %\vspace{-0.5\baselineskip}
        \caption{$p({x_2})$ maximizing ${R_1} + {R_2}$ for $\sqrt {{\rho _1}}  = 1$  and $\sigma  = 0.2.$ (a) $\sqrt {{\rho _2}}  = 2$,(b) $\sqrt {{\rho _2}}  = 4$,(c) $\sqrt {{\rho _2}}  = 100$.}
        \label{fig:3}
\end{figure}

As evident in Fig~\ref{fig:3} when users have different values of peak power constraints, the optimized shape of $p({x_2})$ depends on the value of ${\rho _2}$ and is not necessarily antipodal. As ${\rho _2}$ increases, the resulting $p({x_2})$ converges to a uniform distribution. %Moreover, increasing ${\rho _2}$ enhances the SNR of user 2, hence it decodes before user 1, which is consistent with our decoding strategy.
 \vspace{-1\baselineskip}
 \vspace{-1\baselineskip}

 \vspace{1\baselineskip}
\subsection{Achievable Rate Region in Small Peak Power Constraints}
In this paper the capacity per unit cost is also assessed, that is of interest in the low power regime~\cite{10}, the rates obtained in the sequel with the $0 < \sqrt {{\rho _i}}  \le 1.05$, $i \in \left\{ {1,2} \right\}$ with unit noise variance~\cite{23} which corresponds to small peak power assumption  and are relevant in the desired analyses of Section III. The capacity region of two-user DMC-MAC is achieved using joint and successive decoding in~\cite{43}. In the next Proposition, an achievable rate region is presented for GMAC with small peak power constraints.

\emph{Proposition} 1. \label{prop:1} An achievable rate region for two-user GMAC with small peak power constraints and finite bandwidth using successive decoding is given by
\begin{subequations}\label{eq:18}
\begin{align}
\label{eq:18a}
&{R_1} \le I\left( {{X_1};\left. Y \right|{X_2}} \right) = {\rho _1} - \int\limits_{ - \infty }^\infty  {\frac{{{e^{ - \frac{{{y^2}}}{2}}}}}{{\sqrt {2\pi } }}\log \cosh \left( {{\rho _1} - \sqrt {{\rho _1}} y} \right)dy} , \\ \label{eq:18b}
&{R_2} \le I\left( {{X_2};\left. Y \right|{X_1}} \right){\rm{ = }}{\rho _2} - \int\limits_{ - \infty }^\infty  {\frac{{{e^{ - \frac{{{y^2}}}{2}}}}}{{\sqrt {2\pi } }}\log \cosh \left( {{\rho _2} - \sqrt {{\rho _2}} y} \right)dy} ,\\ \label{eq:18c}
&\begin{array}{l}
{R_1} + {R_2} \le I\left( {{X_1},{X_2};Y} \right){\rm{ = }}{\rho _1} + {\rho _2} - \frac{1}{2}\int\limits_{ - \infty }^\infty  {\frac{{{e^{ - \frac{{{y^2}}}{2}}}}}{{\sqrt {2\pi } }}\log \left( {\cosh \left( {{\rho _1} + {\rho _2} + 2\sqrt {{\rho _1}{\rho _2}}  - \left( {\sqrt {{\rho _1}}  + \sqrt {{\rho _2}} } \right)y} \right)} \right.} \times \\
{\rm{                    }}\left. {{\rm{                                                       }}\cosh \left( {{\rho _1} + {\rho _2} - 2\sqrt {{\rho _1}{\rho _2}}  - \left| {\sqrt {{\rho _1}}  - \sqrt {{\rho _2}} } \right|y} \right)} \right)dy,
\end{array}
\end{align}
\end{subequations}
the input distribution of ${X_i}$ is given by
\begin{equation}\label{eq:19}
p\left( {{x_i}} \right) = \frac{1}{2}\left( {\delta \left( {{x_i} + \sqrt {{\rho _i}} } \right) + \delta \left( {{x_i} - \sqrt {{\rho _i}} } \right)} \right)
\end{equation}
where, $i \in \left\{ {1,2} \right\}$  and $\sqrt {{\rho _i}}  \le 1.05$.

\emph{Proof.} See Appendix A.

In~\eqref{eq:18a}, $\log $ and $\cosh$ are the natural logarithm and hyperbolic cosine functions, respectively. The rates in ~\eqref{eq:18a} are then measured in nats.

The next corollary presents the corner points of the achievable rate region for GMAC with small peak power constraints.

\emph{Corollary} 1. The corner points of the achievable rate region described in Proposition 1 for GMAC with small peak power constraints and finite bandwidth are $\left( {I({X_1};Y),I\left( {{X_2};\left. Y \right|{X_1}} \right)} \right)$  and $\left( {I({X_1};Y),I\left( {{X_2};\left. Y \right|{X_1}} \right)} \right)$, where $I({X_i};Y)$, $i \in \left\{ {1,2} \right\}$, are  given by
\begin{equation}\label{eq:20}
\begin{array}{l}
I\left( {{X_i};Y} \right){\rm{ = }}\\
{\rho _i} - \frac{1}{2}\int\limits_{ - \infty }^\infty  {\frac{{{e^{ - \frac{{{y^2}}}{2}}}}}{{\sqrt {2\pi } }}\log \left( {\frac{{\cosh \left( {{\rho _1} + {\rho _2} + 2\sqrt {{\rho _1}{\rho _2}}  - \left( {\sqrt {{\rho _1}}  + \sqrt {{\rho _2}} } \right)y} \right)\cosh \left( {{\rho _1} + {\rho _2} - 2\sqrt {{\rho _1}{\rho _2}}  - \left| {\sqrt {{\rho _1}}  - \sqrt {{\rho _2}} } \right|y} \right)}}{{{{\left( {\cosh \left( {{\rho _i} - \sqrt {{\rho _i}} y} \right)} \right)}^2}}}} \right)dy} ,
\end{array}
\end{equation}
\emph{Proof}. See Appendix B.

Based on the input distribution in~\eqref{eq:19}, ${\rho _1}$ and ${\rho _2}$  are equal to the signal to noise ratio of users 1 and 2, denoted by $\mathit{SNR_1}$  and $\mathit{SNR_2}$, respectively. As a result, the peak power constraints in fact also imposes an average power constraints. The achievable rate region for ${R_1}$   and ${R_2}$ may now be described in terms of $\mathit{SNR_1}$  and $\mathit{SNR_2}$  and a fixed time sharing parameter $0 \le \alpha  \le 1$  as follows
\begin{subequations}\label{eq:21}
\begin{align}
\label{eq:21a}
&\begin{array}{l}
\bigcup\limits_{0 \le \alpha  \le 1} {\left\{ {{R_1} \le 2\mathit{SNR_1}} \right. - 2\alpha \int\limits_{ - \infty }^\infty  {\frac{{{e^{ - \frac{{{y^2}}}{2}}}}}{{\sqrt {2\pi } }}\log \cosh \left( {\mathit{SNR_1} - \sqrt {\mathit{SNR_1}} y} \right)dy} }  - \left( {1 - \alpha } \right)\int\limits_{ - \infty }^\infty  {\frac{{{e^{ - \frac{{{y^2}}}{2}}}}}{{\sqrt {2\pi } }}} \left\{ {} \right.{\rm{  }}\\
\log \left( {\cosh \left( {\mathit{SNR_1} + \mathit{SNR_2} + 2\sqrt {\mathit{SNR_1}\mathit{SNR_2}}  - \left( {\sqrt {\mathit{SNR_1}}  + \sqrt {\mathit{SNR_2}} } \right)y} \right)} \right) + \\
\log \left( {\cosh \left( {\mathit{SNR_1} + \mathit{SNR_2} - 2\sqrt {\mathit{SNR_1}\mathit{SNR_2}}  - \left| {\sqrt {\mathit{SNR_1}}  - \sqrt {\mathit{SNR_2}} } \right|y} \right)} \right) - \\
\left. {2\log \left( {\cosh \left( {\mathit{SNR_2} - \sqrt {\mathit{SNR_2}} y} \right)} \right)} \right\}dy{\rm{ }}
\end{array} \\ \label{eq:21b}
& \begin{array}{l}
{R_2} \le 2\mathit{SNR_2} - {\rm{ 2}}\left( {1 - \alpha } \right)\int\limits_{ - \infty }^\infty  {\frac{{{e^{ - \frac{{{y^2}}}{2}}}}}{{\sqrt {2\pi } }}\log \cosh \left( {\mathit{SNR_2} - \sqrt {\mathit{SNR_2}} y} \right)dy - } \alpha \int\limits_{ - \infty }^\infty  {\frac{{{e^{ - \frac{{{y^2}}}{2}}}}}{{\sqrt {2\pi } }}} \left\{ {} \right.\\
\log \left( {\cosh \left( {\mathit{SNR_1} + \mathit{SNR_2} + 2\sqrt {\mathit{SNR_1}\mathit{SNR_2}}  - \left( {\sqrt {\mathit{SNR_1}}  + \sqrt {\mathit{SNR_2}} } \right)y} \right)} \right) + \\
\log \left( {\cosh \left( {\mathit{SNR_1} + \mathit{SNR_2} - 2\mathit{SNR_1}\mathit{SNR_2} - \left| {\sqrt {\mathit{SNR_1}}  - \sqrt {\mathit{SNR_2}} } \right|y} \right)} \right) - \\
\left. {\left. {2\log \left( {\cosh \left( {\mathit{SNR_1} - \sqrt {\mathit{SNR_1}} y} \right)} \right)} \right\}dy} \right\}
\end{array}
\end{align}
\end{subequations}
This is obtained using Proposition 1, Corollary 1 and considering a bandwidth of ${W_1} = $ 1 Hz and $2{W_1}$ samples per second.

\begin{figure}[t!]
        \centering
        \includegraphics[width=0.5\textwidth]{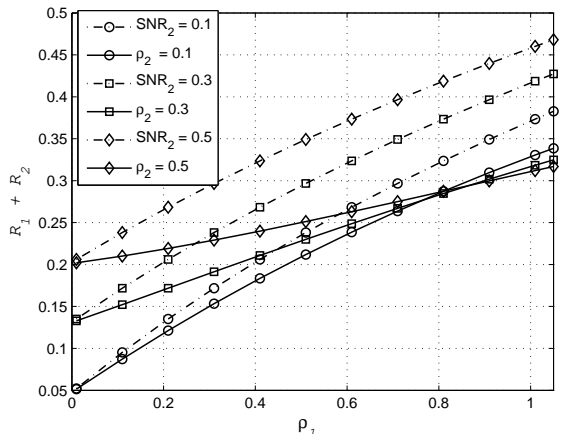}
        \vspace{-1\baselineskip}
        \caption{$R_1 + R_2$  vs. ${\rho _1}$  as a function of ${\rho _2}$  for GMAC (Eq.~\eqref{eq:18c}) with small peak power constraints (dash-dotted curves) or average power constraints (solid curves).}
        \label{fig:4}
\end{figure}

Fig.~\ref{fig:4} shows $R_1 + R_2$ versus ${\rho _1}$ for GMAC (Eq.~\eqref{eq:18a}) with small peak or average power constraints for different values of ${\rho _2}$. The sum rate with small peak power constraints in Proposition 1 is less than that with average power constraints (available in~\cite{43}). Moreover, by increasing the value of peak power constraints the said gap increases.

\begin{figure}[t!]
        \centering
        \includegraphics[width=0.5\textwidth]{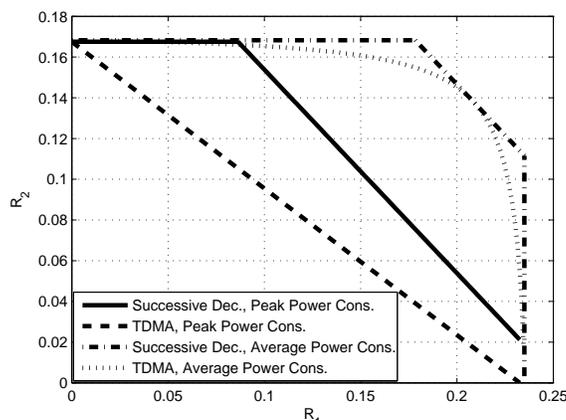}
        \vspace{-1\baselineskip}
        \caption{Achievable rate region of GMAC with peak power or average power constraints and ${\rho _1} = 0.6$, ${\rho _2} = 0.4$.}
        \label{fig:5}
\end{figure}

Another achievable rate region can be obtained using TDMA. It is straight forward to see that the TDMA rates with peak-power constraints and bandwidth of 1 Hz is within the following region
\begin{subequations}\label{eq:22}
\begin{align}
\label{eq:22a}
\bigcup\limits_{0 \le \alpha  \le 1} &{\left\{ {{R_1}} \right. \le 2\alpha \mathit{SNR_1} - 2\alpha \int\limits_{ - \infty }^\infty  {\frac{{{e^{ - \frac{{{y^2}}}{2}}}}}{{\sqrt {2\pi } }}\log \cosh \left( {\mathit{SNR_1} - \sqrt {\mathit{SNR_1}} y} \right)dy} ,} \\\label{eq:22b}
&{R_2} \le 2\left( {1 - \alpha } \right)\mathit{SNR_2} - 2\left( {1 - \alpha } \right)\int\limits_{ - \infty }^\infty  {\frac{{{e^{ - \frac{{{y^2}}}{2}}}}}{{\sqrt {2\pi } }}\log \cosh \left( {\mathit{SNR_2} - \sqrt {\mathit{SNR_2}} y} \right)\left. {dy} \right\}.}
\end{align}
\end{subequations}
Fig.~\ref{fig:5} compares the achievable rate region of GMAC with small peak or average power constraints for successive decoding and TDMA and ${\rho _1} = 0.6$ , ${\rho _2} = 0.4$ . As evident the achievable rate region of successive decoding with peak power constraints is smaller than that with average power constraints for equal average powers. Moreover, only naive TDMA may be used when peak power constraints is applied. Hence, unlike the case with average power constraints, the rate region curves of TDMA and successive decoding with peak power constraints do not touch.

The next proposition examines the capacity of GMAC with small peak power constraints and infinite bandwidth.

\emph{Proposition} 2. The capacity region of GMAC with small peak power constraints and infinite bandwidth is achieved by successive decoding and is given by
\begin{equation}\label{eq:23}
{R_i} \le \frac{{{P_i}}}{{{N_0}}},{\rm{       nat/sec,}} i \in \{1,2\}
\end{equation}
\emph{Proof}. See Appendix C.

The proposition 2 shows that all senders can transmit at their individual capacities, implying that infinite bandwidth removes the interference. In this case, the ratio of the achievable sum rate due to successive decoding and TDMA is given by
\begin{equation}\label{eq:24}
\frac{{{{\left( {{R_1} + {R_2}} \right)}_{{\rm{Succssive Decoding}}}}}}{{{{\left( {{R_1} + {R_2}} \right)}_{{\rm{\textit{TDMA}}}}}}} = \frac{{{P_1} + {P_2}}}{{\alpha {P_1} + \left( {1 - \alpha } \right){P_2}}}
\end{equation}
In the case of GMAC with infinite bandwidth and average power constraints, TDMA is shown to achieve the same performance (capacity)~\cite{43}.

\emph{Remark} 3. The capacity of an AWGN channel with small peak power constraint and infinite bandwidth is given by
\begin{equation}\label{eq:25}
R = \frac{P}{{{N_0}}}{\rm{,       nat/sec}}
\end{equation}
 \vspace{-1\baselineskip}
 \vspace{-1\baselineskip}

\section{Capacity per Unit Cost for AWGN Channel and GMAC with Peak Power Constraints and Finite Bandwidth}
Here, the capacity per unit cost of AWGN channel with small peak power constraint and finite (but large) bandwidth is derived. Next using the results of Proposition 1, the capacity per unit cost region of GMAC with peak power constraints and finite bandwidth is obtained.
 \vspace{-0.5\baselineskip}
\subsection{Capacity Per Unit Cost of AWGN Channel with Peak Power Constraint and Finite Bandwidth}
To obtain the capacity per unit cost of AWGN channel with peak power constraint and finite bandwidth, first the minimum required energy for sending a bit in the channel is to be derived. This is defined as follows~\cite{10}
\begin{equation}\label{eq:26}
{\frac{{{E_b}}}{{{N_0}}}_{\min }} \buildrel \Delta \over = \mathop {\lim }\limits_{\mathit{SNR} \to 0} \frac{{\mathit{SNR}}}{{C\left( {\mathit{SNR}} \right)}},
\end{equation}
where~\eqref{eq:26} is the capacity as a function of $\mathit{SNR}$. The next proposition quantifies~\eqref{eq:26} for the case of our interest.

\emph{Proposition} 3.\label{pro:3} In an AWGN channel with peak power constraint, the minimum energy per bit is given by
\begin{equation}\label{eq:27}
{\frac{{{E_b}}}{{{N_0}}}_{\min }} =  - 1.59{\rm{ dB}}.
\end{equation}
\emph{Proof}. See Appendix D.

This result indicates that ${\frac{{{E_b}}}{{{N_0}}}_{\min }}$ for the AWGN channel with peak power constraint amounts to the same value as that of the AWGN channel with average power constraint~\cite{10}.

Next, we obtain the capacity per unit cost or the slope of spectral efficiency versus ${{{E_b}} \mathord{\left/ {\vphantom {{{E_b}} {{N_0}}}} \right. \kern-\nulldelimiterspace} {{N_0}}}$  curve, ${{{\cal S}}_0}$ in b/s/Hz/3dB at ${\left( {{{{E_b}} \mathord{\left/ {\vphantom {{{E_b}} {{N_0}}}} \right. \kern-\nulldelimiterspace} {{N_0}}}} \right)_{\min }}$. As in this case, the capacity cost function is a concave function of SNR, using Taylor series expansion for a finite (but large) bandwidth, this slope may be quantified as follows~\cite{10}
\begin{subequations}\label{eq:28}
\begin{align}
\label{eq:28a}
{{{\cal S}}_0} & \buildrel \Delta \over =  \mathop {\lim }\limits_{\frac{{{E_b}}}{{{N_0}}} \to {{\frac{{{E_b}}}{{{N_0}}}}_{\min }}} \frac{{C\left( {\frac{{{E_b}}}{{{N_0}}}} \right)}}{{10{{\log }_{10}}\frac{{{E_b}}}{{{N_0}}} - 10{{\log }_{10}}{{\frac{{{E_b}}}{{{N_0}}}}_{\min }}}}10{\log _{10}}2 \\ \label{eq:28b}
&= \frac{{2{{\left[ {\dot C\left( 0 \right)} \right]}^2}}}{{ - \ddot C\left( 0 \right)}}{\rm{.}}
\end{align}
\end{subequations}
where $\dot C\left( 0 \right)$ and $\ddot C\left( 0 \right)$ denote the first and the second derivative of the capacity function, respectively. Thus we have the next Proposition quantifying ${{{\cal S}}_0}$.

\emph{Proposition} 4.\label{pro:4} The capacity per unit cost of AWGN channel with peak power constraint and finite (but large) bandwidth is 1 bit/Joul.

Proof. See Appendix E.
\subsection{Capacity Per Unit Cost of GMAC with Peak Power Constraints and Finite Bandwidth}
Here, the capacity per unit cost of GMAC with peak power constraints and finite bandwidth is derived. We start by deriving the minimum required energy for sending a bit in this channel in the next theorem.

\emph{Theorem} 1. The minimum transmission energy per information bit for GMAC with peak power constraints is achieved by successive decoding and is given by
\begin{equation}\label{eq:29}
{\frac{{{E_1}}}{{{N_0}}}_{\min }} = {\frac{{{E_2}}}{{{N_0}}}_{\min }} =  - 1.59{\rm{ dB}},
\end{equation}
\emph{Proof}. See Appendix F.

\emph{Corollary} 2.\label{coro:2}  When power and rate vanish, the differences between the energies per information bit of a two-user GMAC with peak power constraints using TDMA and those in Theorem 1 are given by
\begin{subequations}\label{eq:30}
\begin{align}
\label{eq:30a}
&\Delta \frac{{{E_1}}}{{{N_0}}} = 10{\log _{10}}{\frac{{{E_1}}}{{{N_0}}}_{\textit{TDMA}}} - 10{\log _{10}}{\frac{{{E_1}}}{{{N_0}}}_{\min }} = 10{\log _{10}}\frac{1}{\alpha }\\
\label{eq:30b}
&\Delta \frac{{{E_2}}}{{{N_0}}} = 10{\log _{10}}{\frac{{{E_2}}}{{{N_0}}}_{\textit{TDMA}}} - 10{\log _{10}}{\frac{{{E_2}}}{{{N_0}}}_{\min }} = 10{\log _{10}}\frac{1}{{1 - \alpha }}
\end{align}
\end{subequations}
in which $0 \le \alpha  \le 1$ is the time sharing parameter for the two users.

\emph{Proof}. See Appendix G.

\emph{Remark} 4.\label{Rem:4} Since $0 \le \alpha  \le 1$, the minimum energies per information bit for GMAC with peak power constraints cannot be achieved by TDMA. This is in contrast to the case with average power constraints, where TDMA is optimum in this sense~\cite{12}.
Due to the antipodal input distribution, imposed by the peak power constraints, the values of peak and average power are equal. As in the problem of interest, the capacity is a concave function of SNR, letting $\frac{{{E_i}}}{{{N_0}}} \to {\frac{{{E_i}}}{{{N_0}}}_{\min }}$ , $i \in \left\{ {1,2} \right\}$  is equivalent to $\mathit{SNR_i} \to 0$~\cite{10}. Hence, the slope region of $S\left( \theta  \right)$ for GMAC is described by the following set of slope pairs
\begin{equation}\label{eq:31}
{{{\cal S}}_i} \buildrel \Delta \over = \mathop {\lim }\limits_{\frac{{{E_i}}}{{{N_0}}} \to {{\frac{{{E_i}}}{{{N_0}}}}_{\min }}} \frac{{{R_i}}}{{10{{\log }_{10}}\frac{{{E_i}}}{{{N_0}}} - 10{{\log }_{10}}{{\frac{{{E_i}}}{{{N_0}}}}_{\min }}}}10{\log _{10}}2
\end{equation}
where, $i \in \left\{ {1,2} \right\}$. Next, we compute a region for achievable rate per unit cost based on the successive decoding strategy. To this end, the following lemma is used.

\emph{Lemma} 2.\label{lem:2} For $\theta  = {{{R_1}} \mathord{\left/ {\vphantom {{{R_1}} {{R_2}}}} \right. \kern-\nulldelimiterspace} {{R_2}}}$, when the powers and rates vanish in two-user GMAC with peak power constraints, we have
\begin{equation}\label{eq:32}
\frac{{{R_1}}}{{{R_2}}} = \frac{{\mathit{SNR_1}}}{{\mathit{SNR_2}}} = \theta .
\end{equation}
\emph{Proof}. As evident in Theorem 1, when the powers and rates vanish, both received energies per bit, ${{{E_i}} \mathord{\left/ {\vphantom {{{E_i}} {{N_0}}}} \right. \kern-\nulldelimiterspace} {{N_0}}},i \in \left\{ {1,2} \right\}$  , approach the same value. Considering ${{{E_i}} \mathord{\left/
 {\vphantom {{{E_i}} {{N_0}}}} \right. \kern-\nulldelimiterspace} {{N_0}}} = {{\mathit{SNR_i}} \mathord{\left/ {\vphantom {{\mathit{SNR_i}} {{R_i}}}} \right.
 \kern-\nulldelimiterspace} {{R_i}}},$ $i \in \left\{ {1,2} \right\}$, the proof is complete.

Similar observation is made for the rate ratio of users in GMAC with average power constraints when powers and rates vanish~\cite{12}.

The next theorem examines the slope region of GMAC with peak power constraints.

\emph{Theorem} 2. For $\theta  = {{{R_1}} \mathord{\left/ {\vphantom {{{R_1}} {{R_2}}}} \right. \kern-\nulldelimiterspace} {{R_2}}}$, an achievable slope region of GMAC with peak power constraints utilizing successive decoding is given by
\begin{equation}\label{eq:33}
S\left( \theta  \right) = \left\{ {\left( {{{{\cal S}}_1},{{{\cal S}}_2}} \right)} \right.:0 \le {{{\cal S}}_1} \le 1,\left. {0 \le {{{\cal S}}_2} \le 1} \right\},
\end{equation}
\emph{Proof}. See Appendix H.

\emph{Proposition} 5.\label{pro:5} Let the rates vanish while $\theta  = {{{R_1}} \mathord{\left/ {\vphantom {{{R_1}} {{R_2}}}} \right.
 \kern-\nulldelimiterspace} {{R_2}}}$ is fixed. The optimum GMAC slope region is achieved by successive decoding as given in~\eqref{eq:33}.

\emph{Proof}. In the slope region of~\eqref{eq:33}, no separate constraints is imposed on the sum rate. As a result, the achieved rate region is a rectangular, where each of its sides is equal to the capacity per unit cost of a single user AWGN channel with small peak power constraints. Hence, the achieved slope region in Theorem 2 is optimum. Therefore, the capacity per unit cost of GMAC with peak power constraints and finite bandwidth is achieved using successive decoding.

In the next proposition, we consider the achievable rate per unit cost of GMAC using TDMA for the case with finite bandwidth and peak power constraints.

\emph{Proposition} 6.\label{pro:6} For $\theta  = {{{R_1}} \mathord{\left/ {\vphantom {{{R_1}} {{R_2}}}} \right. \kern-\nulldelimiterspace} {{R_2}}}$ , the achievable slope region of GMAC with peak power constraints using TDMA is given by
\begin{equation}\label{eq:34}
S\left( \theta  \right) = \left\{ {\left( {{{{\cal S}}_1},{{{\cal S}}_2}} \right):0 \le {{{\cal S}}_1},0 \le {{{\cal S}}_2},{{{\cal S}}_1} + {{{\cal S}}_2} \le 1} \right\}.
\end{equation}
\emph{Proof}. Using Eq.~\eqref{eq:22a} and~\eqref{eq:22b} in~\eqref{eq:31}, we have ${{{\cal S}}_1} = \alpha$  and ${{{\cal S}}_2} = 1 - \alpha$.
It is evident that, regardless of the value of $\theta$ , the slope region achieved by successive decoding is larger than that due to TDMA.

\section{Conclusions}
In this paper, the capacity and the capacity per unit cost of GMAC with power constraints were studied. A numerical approach to characterize the sum rate and the corresponding input distributions in all SNR regimes was proposed. An achievable rate region was analytically derived in low power regimes. Then, the slope region of this achievable rate region, which is due to antipodal signaling and successive decoding, was computed. It was shown that the resulting slope region is in fact optimum and hence identifies the capacity per unit cost region of GMAC with peak power constraints. Table~\ref{Table1} summarizes the transmission strategies over GMAC with different power constraints and with different performance measures, i.e., capacity, capacity per unit cost and minimum transmission energy per information bit.

Future works in this direction include the assessment of capacity per unit cost for other basic channels with peak power constraint. The capacity per unit cost is believed to be the target performance measure in neuronal communications~\cite{6}. An alternate research direction is to investigate the capacity per unit cost for their corresponding channel and transmission models. Theorem 2 shows that from the perspective of capacity per unit cost the transmitters in GMAC could send their data with maximum rate independent of the rate of the other user. Researching an equivalent result in low power and energy efficient neuronal communications is of substantial interest.
\vspace{-0.5\baselineskip}
\begin{table}[h]
\small
\centering
\caption{Summary of transmission strategies over GMAC with different performance measures and power constraints.}
\begin{tabular}{|c|c|c|c|c|c|c|}
\hline
\begin{tabular}[c]{@{}c@{}}Performance \\ Measure\end{tabular}                                       & \begin{tabular}[c]{@{}c@{}}Type of\\  Power \\ Constraint\end{tabular} & Bandwidth                                                       & \begin{tabular}[c]{@{}c@{}}Transceiver \\ Strategy\end{tabular} & Input Distribution                                                          & \begin{tabular}[c]{@{}c@{}}(C)apacity/\\   (A)chievable\\  Rate\end{tabular} & Reference                                                                   \\ \hline
Capacity                                                                                             & Average                                                                & Finite                                                          & \begin{tabular}[c]{@{}c@{}}Successive\\  decoding\end{tabular}  & Gaussian                                                                    & C                                                                            & \cite{43}                                                                   \\ \hline
Capacity                                                                                             & Average                                                                & Infinite                                                        & TDMA                                                            & Gaussian                                                                    & C                                                                            & \cite{43}                                                                   \\ \hline
Capacity                                                                                             & Peak                                                                   & \begin{tabular}[c]{@{}c@{}}Finite \\   (but large)\end{tabular} & \begin{tabular}[c]{@{}c@{}}Successive\\  decoding\end{tabular}  & \begin{tabular}[c]{@{}c@{}}Equiprobable\\  antipodal signaling\end{tabular} & A                                                                            & Proposition 1                                                               \\ \hline
Capacity                                                                                             & Peak                                                                   & Infinite                                                        & \begin{tabular}[c]{@{}c@{}}Successive \\ decoding\end{tabular}  & \begin{tabular}[c]{@{}c@{}}Equiprobable\\  antipodal signaling\end{tabular} & C                                                                            & Proposition 2                                                               \\ \hline
\begin{tabular}[c]{@{}c@{}}Capacity \\   per unit cost\end{tabular}                                  & Average                                                                & \begin{tabular}[c]{@{}c@{}}Finite \\   (but large)\end{tabular} & \begin{tabular}[c]{@{}c@{}}Successive \\ decoding\end{tabular}  & Gaussian                                                                    & C                                                                            & \cite{12}                                                                   \\ \hline
\begin{tabular}[c]{@{}c@{}}Capacity \\   per unit cost\end{tabular}                                  & Average                                                                & Infinite                                                        & TDMA                                                            & Gaussian                                                                    & C                                                                            & \cite{5}                                                                    \\ \hline
\begin{tabular}[c]{@{}c@{}}Capacity \\   per unit cost\end{tabular}                                  & Peak                                                                   & \begin{tabular}[c]{@{}c@{}}Finite \\   (but large)\end{tabular} & \begin{tabular}[c]{@{}c@{}}Successive\\  decoding\end{tabular}  & \begin{tabular}[c]{@{}c@{}}Equiprobable\\  antipodal signaling\end{tabular} & C                                                                            & Theorem 2                                                                   \\ \hline
\begin{tabular}[c]{@{}c@{}}Capacity \\   per unit cost\end{tabular}                                  & Peak                                                                   & Infinite                                                        & \begin{tabular}[c]{@{}c@{}}Successive\\  decoding\end{tabular}  & \begin{tabular}[c]{@{}c@{}}Equiprobable\\  antipodal signaling\end{tabular} & C                                                                            & \begin{tabular}[c]{@{}c@{}}Direct result of \\   Proposition 2\end{tabular} \\ \hline
\begin{tabular}[c]{@{}c@{}}Minimizing \\ transmission\\   energy per\\  information bit\end{tabular} & Average                                                                & Finite/Infinite                                                 & TDMA                                                            & Gaussian                                                                    & -                                                                            & \cite{12}                                                                   \\ \hline
\begin{tabular}[c]{@{}c@{}}Minimizing\\  transmission \\ energy per \\ information bit\end{tabular}  & Peak                                                                   & Finite/Infinite                                                 & \begin{tabular}[c]{@{}c@{}}Successive\\  decoding\end{tabular}  & \begin{tabular}[c]{@{}c@{}}Equiprobable \\ antipodal signaling\end{tabular} & -                                                                            & Theorem 1                                                                   \\ \hline
\end{tabular}
\label{Table1}
\end{table}

\section{Appendices}
\subsection{Proof of Proposition 1}
Noting~\eqref{eq:18a}, the rate ${R_1}$ is constrained by
\begin{equation}\label{eq:35}
\begin{array}{l}
{R_1} \le I\left( {{X_1};\left. Y \right|{X_2}} \right) = h\left( {\left. {{X_1} + {X_2} + Z} \right|{X_2}} \right) - h\left( {\left. {{X_1} + {X_2} + Z} \right|{X_1},{X_2}} \right)\\
{\rm{                           }} = h\left( {{X_1} + Z} \right) - h\left( Z \right)\\
{\rm{                           }} = h\left( {{X_1} + Z} \right) - 0.5\log 2\pi e,{\rm{  }}
\end{array}
\end{equation}
The above rate using bipolar distribution as,
\begin{equation}\label{eq:38}
p\left( x_1 \right) = \frac{1}{2}\left( {\delta \left( {x_1 + \sqrt \rho_1  } \right) + \delta \left( {x_1 - \sqrt \rho_1  } \right)} \right).
\end{equation}
is given by~\cite{23}
%\end{subequations}
\begin{equation}~\label{eq:39}
R_1 \le \rho_1  - \int\limits_{ - \infty }^\infty  {\frac{{{e^{ - \frac{{{y^2}}}{2}}}}}{{\sqrt {2\pi } }}\log \cosh \left( {\rho_1  - \sqrt \rho_1  y} \right)dy.}
\end{equation}
The sum rate of ${R_1} + {R_2}$ is limited by
\begin{equation}\label{eq:40}
\begin{array}{l}
{R_1} + {R_2} \le I\left( {{X_1},{X_2};Y} \right)\\
{\rm{           }} = h\left( {{X_1} + {X_2} + Z} \right) - h\left( Z \right)\\
{\rm{           }} = h\left( {U + Z} \right) - \frac{1}{2}\log 2\pi e,{\rm{ }}
\end{array}
\end{equation}
The random variables $X_1$ and $X_2$ are real-valued and independent in MAC, hence $p\left( U \right)$  is obtained by the convolution of $p\left( {{X_1}} \right)$  and $p\left( {{X_2}} \right)$ . Using (19) and with some manipulations, $p\left( U \right)$  is given by
\begin{equation}\label{eq:42}
\begin{array}{l}
p(U) = \frac{1}{4}\left[ {\delta \left( {U - \sqrt {{\rho _1}}  - \sqrt {{\rho _2}} } \right) + } \right.\delta \left( {U - \sqrt {{\rho _1}}  + \sqrt {{\rho _2}} } \right) + \\
{\rm{                 }}\delta \left( {U + \sqrt {{\rho _1}}  - \sqrt {{\rho _2}} } \right) + \left. {\delta \left( {U + \sqrt {{\rho _1}}  + \sqrt {{\rho _2}} } \right)} \right].
\end{array}
\end{equation}
It is obvious that, $p\left( U \right)$ in~\eqref{eq:42} is not an antipodal distribution. The uniqueness of capacity achieving distribution for AWGN channel with peak power constraint is proven in~\cite{22}. Hence, there is not any distribution for random variables ${X_1}$ and ${X_2}$, which jointly maximizes $I\left( {Y;{X_1},{X_2}} \right)$, $I\left( {{X_1};\left. Y \right|{X_2}} \right)$  and $I\left( {{X_2};\left. Y \right|{X_1}} \right)$. Therefore, with input distribution of~\eqref{eq:19} which maximizes~\eqref{eq:18a} and~\eqref{eq:18b}, an inner bound for ~\eqref{eq:18c} is derived as
\begin{subequations}\label{eq:43}
\begin{align}
\label{eq:43a}
\nonumber
{R_1} + {R_2} \le
& \frac{{\rm{1}}}{{\rm{4}}}h\left( {\left. {\sqrt {{\rho _1}}  + \sqrt {{\rho _2}}  + Z} \right|{X_1} = \sqrt {{\rho _1}} ,{X_2} = \sqrt {{\rho _2}} } \right) + \\ \nonumber &\frac{{\rm{1}}}{{\rm{4}}}h\left( {\left. { - \sqrt {{\rho _1}}  - \sqrt {{\rho _2}}  + Z} \right|{X_1} =  - \sqrt {{\rho _1}} ,{X_2} =  - \sqrt {{\rho _2}} } \right) + \\ \nonumber
&\frac{{\rm{1}}}{{\rm{4}}}h\left( {\left. {\sqrt {{\rho _1}}  - \sqrt {{\rho _2}}  + Z} \right|{X_1} = \sqrt {{\rho _1}} ,{X_2} = \sqrt {{\rho _2}} } \right) +  \\  & \frac{{\rm{1}}}{{\rm{4}}}h\left( {\left. {\sqrt {{\rho _2}}  - \sqrt {{\rho _1}}  + Z} \right|{X_1} =  - \sqrt {{\rho _1}} ,{X_2} =  - \sqrt {{\rho _2}} } \right) - \frac{1}{2}\log 2\pi e \\ \nonumber
 = & \frac{{\rm{1}}}{{\rm{2}}}\left\{ {\frac{1}{2}\log 2\pi e + {\rho _1} + {\rho _2} + 2\sqrt {{\rho _1}{\rho _2}}  - } \right.\\ \nonumber
 &\left. {\int\limits_{ - \infty }^\infty  {\frac{{{e^{ - \frac{{{y^2}}}{2}}}}}{{\sqrt {2\pi } }}\log \cosh \left( {{\rho _1} + {\rho _2} + 2\sqrt {{\rho _1}{\rho _2}}  - \left( {\sqrt {{\rho _1}}  + \sqrt {{\rho _2}} } \right)y} \right)dy} } \right\} + \\ \nonumber
 & \frac{{\rm{1}}}{{\rm{2}}}\left\{ {\frac{1}{2}\log 2\pi e + {\rho _1} + {\rho _2} - 2\sqrt {{\rho _1}{\rho _2}}  - } \right.\\ \label{eq:43b}
 & \left. {\int\limits_{ - \infty }^\infty  {\frac{{{e^{ - \frac{{{y^2}}}{2}}}}}{{\sqrt {2\pi } }}\log \cosh \left( {{\rho _1} + {\rho _2} - 2\sqrt {{\rho _1}{\rho _2}}  - \left| {\sqrt {{\rho _1}}  - \sqrt {{\rho _2}} } \right|y} \right)dy} } \right\} - \frac{1}{2}\log 2\pi e
\end{align}
\end{subequations}
\eqref{eq:43b} is derived from~\eqref{eq:43a} using~\eqref{eq:38}, by some simple manipulation~\eqref{eq:18c} is derived from~\eqref{eq:43b}.
 \vspace{-1.5\baselineskip}
\subsection{Proof of Corollary 1}
A corner point of the achievable rate region in Proposition 1 is $\left( {I({X_1};Y),I\left( {{X_2};\left. Y \right|{X_1}} \right)} \right)$, where $I\left( {{X_2};\left. Y \right|{X_1}} \right)$  is calculated in~\eqref{eq:18a} and $I({X_1};Y)$ derived as
\begin{subequations}\label{eq:44}
\begin{align}
\label{eq:44a}
\nonumber I\left( {{X_1};Y} \right) & = h\left( {{X_1} + {X_2} + Z} \right) - h\left( {{X_2} + Z} \right) = \frac{1}{2}\log 2\pi e + {\rho _1} + {\rho _2}\\ \nonumber
& - \frac{1}{2}\int\limits_{ - \infty }^\infty  {\frac{{{e^{ - \frac{{{y^2}}}{2}}}}}{{\sqrt {2\pi } }}\log \left( {\cosh \left( {{\rho _1} + {\rho _2} + 2\sqrt {{\rho _1}{\rho _2}}  - \left( {\sqrt {{\rho _1}}  + \sqrt {{\rho _2}} } \right)y} \right)} \right.}  \times \\ \nonumber
& \left. {\cosh \left( {{\rho _1} + {\rho _2} - 2\sqrt {{\rho _1}{\rho _2}}  - \left| {\sqrt {{\rho _1}}  - \sqrt {{\rho _2}} } \right|y} \right)} \right)dy  \\ & - \frac{1}{2}\log 2\pi e - {\rho _2} + \int\limits_{ - \infty }^\infty  {\frac{{{e^{ - \frac{{{y^2}}}{2}}}}}{{\sqrt {2\pi } }}\log \cosh \left( {{\rho _2} - \sqrt {{\rho _2}} y} \right)dy}  = {\rho _1} - \frac{1}{2}\int\limits_{ - \infty }^\infty  {\frac{{{e^{ - \frac{{{y^2}}}{2}}}}}{{\sqrt {2\pi } }}} \times \\ \label{eq:44b}
\log &\left( {\frac{{\cosh \left( {{\rho _1} + {\rho _2} + 2\sqrt {{\rho _1}{\rho _2}}  - \left( {\sqrt {{\rho _1}}  + \sqrt {{\rho _2}} } \right)y} \right)\cosh \left( {{\rho _1} + {\rho _2} - 2\sqrt {{\rho _1}{\rho _2}}  - \left| {\sqrt {{\rho _1}}  - \sqrt {{\rho _2}} } \right|y} \right)}}{{{{\left[ {\cosh \left( {{\rho _2} - \sqrt {{\rho _2}} y} \right)} \right]}^2}}}} \right)dy
\end{align}
\end{subequations}
Equation~\eqref{eq:44b} follows from~\eqref{eq:44a} using~\eqref{eq:38}. The other corner point is obtained similarly.
\vspace{-1\baselineskip}
\subsection{Proof of Proposition 2}
Noting~\eqref{eq:19}, we replace ${\rho _1}$ in~\eqref{eq:22a} with $\mathit{SNR_1}$, and obtain the following
\begin{equation}\label{eq:45}
{R_1} \le \mathit{SNR_1} - \int\limits_{ - \infty }^\infty  {\frac{{{e^{ - \frac{{{y^2}}}{2}}}}}{{\sqrt {2\pi } }}\log \cosh \left( {\mathit{SNR_1} - \sqrt {\mathit{SNR_1}} y} \right)dy} ,
\end{equation}
If the channel bandwidth of user 1 is ${W_1}$, since there are $2{W_1}$ samples per second, the upper bound of ${R_1}$. Also, by replacing $\mathit{SNR_1}$  with ${{{P_1}} \mathord{\left/ {\vphantom {{{P_1}} {\left( {{N_0}{W_1}} \right)}}} \right. \kern-\nulldelimiterspace} {\left( {{N_0}{W_1}} \right)}}$   and as ${W_1}$ approaches infinity,~\eqref{eq:45} can be rewritten as
\begin{equation}\label{eq:46}
{R_1} \le \mathop {\lim }\limits_{{W_1} \to \infty } 2{W_1}\left( {\frac{P_1}{{{N_0}{W_1}}} - \int\limits_{ - \infty }^\infty  {\frac{{{e^{ - \frac{{{y^2}}}{2}}}}}{{\sqrt {2\pi } }}\log \cosh \left( {\frac{P_1}{{{N_0}{W_1}}} - \sqrt {\frac{P_1}{{{N_0}{W_1}}}} y} \right)dy} } \right).
\end{equation}
As ${W_1}$ approaches infinity, $\chi  = \frac{P_1}{{{N_0}{W_1}}} - \sqrt {\frac{P_1}{{{N_0}{W_1}}}} y$  tends to zero. Moreover, the function of $\log \cosh \left( . \right)$ is infinitely differentiable in the neighborhood of zero. Hence, $\log \cosh \left( x \right)$  can be replaced by its following Taylor expansion
\begin{equation}\label{eq:47}
\log \cosh \left( \chi  \right) = \frac{{{\chi ^2}}}{2} - \frac{{{\chi ^4}}}{{12}} + \frac{{{\chi ^6}}}{{45}} - \frac{{17{\chi ^8}}}{{2520}} + ...
\end{equation}
Hence, ${R_1}$ is bounded by
\begin{equation}\label{eq:48}
\begin{array}{l}
{R_1} \le \mathop {\lim }\limits_{{W_1} \to \infty } \\
2{W_1}\left( {\frac{{{P_1}}}{{{N_0}{W_1}}} - \frac{1}{2}\int\limits_{ - \infty }^\infty  {\frac{{{e^{ - \frac{{{y^2}}}{2}}}}}{{\sqrt {2\pi } }}\left[ {\frac{1}{2}{{\left( {\frac{{{P_1}}}{{{N_0}{W_1}}} - \sqrt {\frac{P_1}{{{N_0}{W_1}}}} y} \right)}^2} - \frac{1}{{12}}{{\left( {\frac{{{P_1}}}{{{N_0}{W_1}}} - \sqrt {\frac{{{P_1}}}{{{N_0}{W_1}}}} y} \right)}^4} + ...} \right]dy} } \right)\\
{\rm{    }} = \mathop {\lim }\limits_{{W_1} \to \infty } 2{W_1}\left( {\frac{{{P_1}}}{{2{N_0}{W_1}}} - \frac{1}{2}{{\left( {\frac{{{P_1}}}{{{N_0}{W_1}}}} \right)}^2}} \right) = \frac{{{P_1}}}{{{N_0}}}.
\end{array}
\end{equation}
The upper bound of ${R_2}$  in~\eqref{eq:23} is derived in a similar manner. In the same direction, the upper bound of ${R_1} + {R_2}$  when $W$ tends to infinity can be written as
\begin{eqnarray}\label{eq:49}
\nonumber
{R_1} + {R_2} \le \frac{{2\left( {{P_1} + {P_2}} \right)}}{{{N_0}}} - \mathop {{\rm{lim}}}\limits_{W \to \infty } W\left\{ {\int\limits_{ - \infty }^\infty  {\frac{{{e^{ - \frac{{{y^2}}}{2}}}}}{{\sqrt {2\pi } }}\log \cosh \left( {\frac{{{{\left( {\sqrt {{P_1}}  + \sqrt {{P_2}} } \right)}^2}}}{{{N_0}W}} - \frac{{\sqrt {{P_1}}  + \sqrt {{P_2}} }}{{\sqrt {{N_0}W} }}y} \right)} dy} \right. \\
\left. {{\rm{                                               }}\int\limits_{ - \infty }^\infty  {\frac{{{e^{ - \frac{{{y^2}}}{2}}}}}{{\sqrt {2\pi } }}\log } \cosh \left( {\frac{{{{\left( {\sqrt {{P_1}}  - \sqrt {{P_2}} } \right)}^2}}}{{{N_0}W}} - \frac{{\left| {\sqrt {{P_1}}  - \sqrt {{P_2}} } \right|}}{{\sqrt {{N_0}W} }}y} \right)dy} \right\},
\end{eqnarray}
Using the Taylor expansion of $\log (\cosh (\chi ))$ in~\eqref{eq:47}, we have
\begin{equation}\label{eq:50}
\begin{array}{l}
{R_1} + {R_2} \le \\
\mathop {\lim }\limits_{W \to \infty } 2W\left\{ {\frac{{{P_1} + {P_2}}}{{{N_0}W}}} \right. - \frac{1}{2}\left[ {\frac{1}{2}\frac{{{{\left( {\sqrt {{P_1}}  + \sqrt {{P_2}} } \right)}^2}}}{{{N_0}W}} + \frac{1}{2}\frac{{{{\left( {\sqrt {{P_1}}  + \sqrt {{P_2}} } \right)}^4}}}{{N_0^2{W^2}}}} \right.\left. {\left. { + \frac{1}{2}\frac{{{{\left( {\sqrt {{P_1}}  - \sqrt {{P_2}} } \right)}^2}}}{{{N_0}W}} + \frac{1}{2}\frac{{{{\left( {\sqrt {{P_1}}  - \sqrt {{P_2}} } \right)}^4}}}{{N_0^2{W^2}}}} \right]} \right\} = \\
\mathop {\lim }\limits_{W \to \infty } 2W\left\{ {\frac{{{P_1} + {P_2}}}{{{N_0}W}} - \frac{1}{2}\left. {\left[ {\frac{{{P_1} + {P_2}}}{{{N_0}W}} + \frac{1}{2}\frac{{{{\left( {\sqrt {{P_1}}  + \sqrt {{P_2}} } \right)}^4} + {{\left( {\sqrt {{P_1}}  - \sqrt {{P_2}} } \right)}^4}}}{{N_0^2{W^2}}}} \right.} \right]} \right\}{\rm{ = }}\frac{{{P_1}}}{{{N_0}}} + \frac{{{P_2}}}{{{N_0}}}.
\end{array}
\end{equation}
The achievable rate region of GMAC with peak power constraints and infinite bandwidth is a rectangular region. As a result, the constraints on sum rate will be inactive and the antipodal input distribution for transmitted signal of both users maximizes both ${R_1}$ and ${R_2}$. Hence, the achievable rate region in~\eqref{eq:23} is the capacity region.

\subsection{Proof of Proposition 3}
Noting~\eqref{eq:19}, we replace $\rho_1$ in~\eqref{eq:39} with $\mathit{SNR}$, and since there are $2W$ samples per second, if  $2W$ = 1 Hz, $C\left( \mathit{SNR} \right)$  is given by
\begin{equation}\label{eq:51}
C\left( \mathit{SNR} \right) = 2\mathit{SNR} - 2\int_{ - \infty }^\infty  {\frac{{{e^{ - {{{y^2}} \mathord{\left/
 {\vphantom {{{y^2}} 2}} \right.
 \kern-\nulldelimiterspace} 2}}}}}{{\sqrt {2\pi } }}} \log \cosh \left( {\mathit{SNR} - \sqrt {\mathit{SNR}} y} \right)dy.
\end{equation}
Using~\eqref{eq:26} and~\eqref{eq:51} for $C\left( \mathit{SNR} \right)$, ${\frac{{{E_n}}}{{{N_0}}}_{\min }}$ , which denotes  the minimum required energy for sending a nat, is derived
\begin{equation}\label{eq:52}
{\frac{{{E_n}}}{{{N_0}}}_{\min }} = \mathop {\lim }\limits_{\mathit{SNR} \to 0} \frac{{\mathit{SNR}}}{{2\mathit{SNR} - 2\int_{ - \infty }^\infty  {\frac{{{e^{ - {{{y^2}} \mathord{\left/
 {\vphantom {{{y^2}} 2}} \right.
 \kern-\nulldelimiterspace} 2}}}}}{{\sqrt {2\pi } }}} \log \cosh \left( {\mathit{SNR} - \sqrt \mathit{SNR} y} \right)dy}}
\end{equation}
Noting~\eqref{eq:47}, we have
 \begin{subequations}\label{eq:53}
 \begin{align}
 \label{eq:53a}
{\frac{{{E_n}}}{{{N_0}}}_{\min }} =& \mathop {\lim }\limits_{\mathit{SNR} \to 0} \frac{{\mathit{SNR}}}{{2\mathit{SNR} - \int_{ - \infty }^\infty  {\frac{{{e^{ - {{{y^2}} \mathord{\left/
 {\vphantom {{{y^2}} 2}} \right.
 \kern-\nulldelimiterspace} 2}}}}}{{\sqrt {2\pi } }}} {{\left( {\mathit{SNR} - \sqrt {\mathit{SNR}} y} \right)}^2}dy}}\\ \label{eq:53b}
   = &\mathop {\lim }\limits_{\mathit{SNR} \to 0} \frac{{\mathit{SNR}}}{{\mathit{SNR} - {\mathit{SNR^2}}}} = 1.
 \end{align}
 \end{subequations}
 Translating the results to Joules/bit we have,
 \begin{equation}\label{eq:54}
 {\frac{{{E_b}}}{{{N_0}}}_{\min }} = \frac{1}{{\log _2^e}} = 0.6931 =  - 1.59{\rm{ dB}}
 \end{equation}

\subsection{Proof of Proposition 4}
Using~\eqref{eq:28b}, we need to calculate $\dot C\left(  \cdot  \right)$  and $\ddot C\left(  \cdot  \right)$ , where $C\left( {\frac{{{E_b}}}{{{N_0}}}} \right)$  is equal to $C\left( \mathit{SNR} \right)$  in~\eqref{eq:51}. Noting~\eqref{eq:47} and with some manipulations we have
\begin{equation}\label{eq:55}
C\left( {\frac{{{E_n}}}{{{N_0}}}} \right) \simeq \frac{{{E_n}}}{{{N_0}}} - {\left( {\frac{{{E_n}}}{{{N_0}}}} \right)^2}.
\end{equation}
Thus, using~\eqref{eq:28b} and $\dot C\left( 0 \right)$  and $\ddot C\left( 0 \right)$  for this case amounts to 1 and -2, respectively. Hence, using~\eqref{eq:55} we have ${{{\cal S}}_0} = 1$.

\subsection{Proof of Theorem 1}
Consider a fixed time-sharing parameter $0 \le \alpha  \le 1$. Using~\eqref{eq:21a} in~\eqref{eq:26}, we obtain
\begin{equation}\label{eq:56}
\begin{array}{l}
{\frac{{{E_1}}}{{{N_0}}}_{\min }} = \mathop {\lim }\limits_{\mathit{SNR_1} \to 0} \frac{{\mathit{SNR_1}}}{{2\mathit{SNR_1} - 2\alpha \int\limits_{ - \infty }^\infty  {\frac{{{e^{ - \frac{{{y^2}}}{2}}}}}{{\sqrt {2\pi } }}\log \cosh \left( {\mathit{SNR_1} - \sqrt {\mathit{SNR_1}} y} \right)dy}  - \left( {1 - \alpha } \right)\int\limits_{ - \infty }^\infty  {\frac{{{e^{ - \frac{{{y^2}}}{2}}}}}{{\sqrt {2\pi } }}}  \times }}\\
\frac{{}}{{\left\{ {\log \left[ {\cosh \left( {\mathit{SNR_1} + \mathit{SNR_2} + 2\sqrt {\mathit{SNR_1}\mathit{SNR_2}}  - \left( {\sqrt {\mathit{SNR_1}}  + \sqrt {\mathit{SNR_2}} } \right)y} \right)} \right]} \right. + }}\\
\frac{{}}{{\left. {\log \left[ {\cosh \left( {\mathit{SNR_1} + \mathit{SNR_2} - 2\sqrt {\mathit{SNR_1}\mathit{SNR_2}}  - \left| {\sqrt {\mathit{SNR_1}}  - \sqrt {\mathit{SNR_2}} } \right|y} \right)} \right] - 2\log \left[ {\cosh \left( {\mathit{SNR_2} - \sqrt {\mathit{SNR_2}} y} \right)} \right]} \right\}}}
\end{array}
\end{equation}
Noting~\eqref{eq:56}, we have
\begin{equation}\label{eq:57}
\begin{array}{l}
{\frac{{{E_1}}}{{{N_0}}}_{\min }} = \mathop {\lim }\limits_{\mathit{SNR_1} \to 0} \frac{{\mathit{SNR_1}}}{{{R_1}}} = \mathop {\lim }\limits_{\mathit{SNR_1} \to 0} \frac{{\mathit{SNR_1}}}{{2\mathit{SNR_1} - \alpha \left( {\mathit{SNR_1} + \mathit{SNR_1^2}} \right) - \frac{{\left( {1 - \alpha } \right)}}{2}\left[ {{{\left( {\sqrt {\mathit{SNR_1}}  + \sqrt {\mathit{SNR_2}} } \right)}^2} + {{\left( {\sqrt {\mathit{SNR_1}}  + \sqrt {\mathit{SNR_2}} } \right)}^4} + } \right.}}\\
\frac{{}}{{\left. {{{\left( {\sqrt {\mathit{SNR_1}}  - \sqrt {\mathit{SNR_2}} } \right)}^2} + {{\left( {\sqrt {\mathit{SNR_1}}  - \sqrt {\mathit{SNR_2}} } \right)}^4} - 2\left( {\mathit{SNR_2} + \mathit{SNR_2^2}} \right)} \right]}} = \\
\mathop {\lim }\limits_{\mathit{SNR_1} \to 0} \frac{{\mathit{SNR_1}}}{{2\mathit{SNR_1} - \alpha \left( {\mathit{SNR_1} + \mathit{SNR_1^2}} \right) - \frac{{\left( {1 - \alpha } \right)}}{2}\left[ {\left. {2\mathit{SNR_1} + 2\mathit{SNR_1^2} + 12\mathit{SNR_1}\mathit{SNR_2}} \right]} \right.}} = \\
\mathop {\lim }\limits_{\mathit{SNR_1} \to 0} \frac{{\mathit{SNR_1}}}{{\mathit{SNR_1} - \mathit{SNR_1^2} - 6\left( {1 - \alpha } \right)\mathit{SNR_1}\mathit{SNR_2}}} = \mathop {\lim }\limits_{\mathit{SNR_1} \to 0} {\left. {\frac{1}{{1 - \mathit{SNR_1} - 6\left( {1 - \alpha } \right)\mathit{SNR_2}}}} \right|_{\mathit{SNR_2} = 0}} = 1
\end{array}
\end{equation}
Translating the results to joules/bit we have,
\begin{equation}\label{eq:58}
{\frac{{{E_1}}}{{{N_0}}}_{\min }} = \frac{1}{{\log _2^e}} = 0.6931 =  - 1.59{\rm{ dB}}
\end{equation}
Following similar steps, we obtain similar results for user 2. The above analyses show that for communication over a GMAC one can achieve the ultimate single user performance (see Proposition 4) with successive decoding.
\subsection{Proof of Corollary 2}
Consider a fixed time-sharing parameter $0 < \alpha  < 1$. Using~\eqref{eq:22a} and~\eqref{eq:22b} in~\eqref{eq:26}, we have
\begin{subequations}\label{eq:59}
\begin{align}
\label{eq:59a}
{\frac{{{E_1}}}{{{N_0}}}_{\textit{TDMA}}}& = \mathop {\lim }\limits_{\mathit{SNR_1} \to 0} \frac{{\mathit{SNR_1}\log _e^2}}{{2\alpha \mathit{SNR_1} - 2\alpha \int\limits_{ - \infty }^\infty  {\frac{{{e^{ - \frac{{{y^2}}}{2}}}}}{{\sqrt {2\pi } }}\log \cosh \left( {\mathit{SNR_1} - \sqrt {\mathit{SNR_1}} y} \right)dy} }},\\
\label{eq:59b}
{\frac{{{E_2}}}{{{N_0}}}_{\textit{TDMA}}}& = \mathop {\lim }\limits_{\mathit{SNR_2} \to 0} \frac{{\mathit{SNR_2}\log _e^2}}{{2\left( {1 - \alpha } \right)\mathit{SNR_2} - 2\left( {1 - \alpha } \right)\int\limits_{ - \infty }^\infty  {\frac{{{e^{ - \frac{{{y^2}}}{2}}}}}{{\sqrt {2\pi } }}\log \cosh \left( {\mathit{SNR_2} - \sqrt {\mathit{SNR_2}} y} \right)dy} }},
\end{align}
\end{subequations}
Noting~\eqref{eq:47} and using Hospital’s rule, we have
\begin{subequations}\label{eq:60}
\begin{align}
\label{eq:60a}
{\frac{{{E_1}}}{{{N_0}}}_{\textit{TDMA}}} & = \frac{{\log _e^2}}{\alpha },\\
\label{eq:60b}
{\frac{{{E_2}}}{{{N_0}}}_{\textit{TDMA}}} & = \frac{{\log _e^2}}{{1 - \alpha }},
\end{align}
\end{subequations}
Considering $0 \le \alpha  \le 1$ , we note that ${\frac{{{E_i}}}{{{N_0}}}_{\textit{TDMA}}} \ge {\log _e}2$, $i \in \left\{ {1,2} \right\}$. Hence, the minimum energy per nat is not achievable using TDMA. Converting nat to bit, the difference between ${\frac{{{E_i}}}{{{N_0}}}_{\textit{TDMA}}}$ and ${\frac{{{E_i}}}{{{N_0}}}_{\min }}$   in dB, $\Delta \frac{{{E_i}}}{{{N_0}}}$  , is given by~\eqref{eq:30a} and~\eqref{eq:30b}. As stated, the peak power constraints imposes the input distribution in~\eqref{eq:19}, and hence only naive TDMA may be utilized.

\subsection{Proof of Theorem 2}
Considering~\eqref{eq:21a} and~\eqref{eq:21b} and noting~\eqref{eq:47}, we have
\begin{equation}\label{eq:61}
\begin{array}{l}
\bigcup\limits_{0 \le \alpha  \le 1} {\left\{ {{R_i} \le \alpha \left( {\mathit{SNR_i} - \mathit{SNR_i^2} + {{\cal O}}\left( \mathit{SNR_i^4} \right)} \right)} \right.}  + \left( {1 - \alpha } \right)\mathit{SNR_i}\\
{\rm{             }} + \frac{1}{2}\left( {1 - \alpha } \right)\left( {\left( {1 + \frac{1}{\theta } + \frac{2}{{\sqrt \theta  }}} \right)\mathit{SNR_i} - {{\left( {1 + \frac{1}{\theta } + \frac{2}{{\sqrt \theta  }}} \right)}^2}\mathit{SNR_i^2} + {{\cal O}}\left( \mathit{SNR_i^4} \right)} \right)\\
{\rm{             }} - \frac{1}{2}\left( {1 - \alpha } \right)\left( {1 + \frac{1}{\theta } + \frac{2}{{\sqrt \theta  }}} \right)\mathit{SNR_i}\\
{\rm{            }} + \frac{1}{2}\left( {1 - \alpha } \right)\left( {\left( {1 + \frac{1}{\theta } - \frac{2}{{\sqrt \theta  }}} \right)\mathit{SNR_i} - {{\left( {1 + \frac{1}{\theta } - \frac{2}{{\sqrt \theta  }}} \right)}^2}\mathit{SNR_i^2} + {{\cal O}}\left( \mathit{SNR_i^4} \right)} \right)\\
{\rm{            }} - \frac{1}{2}\left( {1 - \alpha } \right)\left( {1 + \frac{1}{\theta } - \frac{2}{{\sqrt \theta  }}} \right)\mathit{SNR_i}\\
{\rm{            }} - \left( {1 - \alpha } \right)\left( {\left( {\frac{1}{\theta }} \right)\mathit{SNR_i} - {{\left( {\frac{1}{\theta }} \right)}^2}\mathit{SNR_i^2} + {{\cal O}}\left( \mathit{SNR_i^4} \right)} \right) + \left( {1 - \alpha } \right)\left( {\frac{1}{\theta }} \right)\left. {\mathit{SNR_i}} \right\},
\end{array}
\end{equation}
where, $i \in \left\{ {1,2} \right\}$. Representing the RHS of~\eqref{eq:61} by $R_i^u$, and using~\eqref{eq:31}, and~\eqref{eq:28b}, we have
\begin{equation}\label{eq:62}
\begin{array}{l}
{{{\cal S}}_i} = \mathop {\lim }\limits_{\frac{{{E_i}}}{{{N_0}}} \to {{\frac{{{E_i}}}{{{N_0}}}}_{\min }}} \frac{{R_i^u}}{{10{{\log }_{10}}\mathit{SNR} - 10{{\log }_{10}}\mathit{SNR_{\min }}}}10{\log _{10}}2 = \mathop {\lim }\limits_{\mathit{SNR_i} \to \mathit{SNR_{\min }}} \frac{{ - 2{{\left[ {\dot R_i^u\left( 0 \right)} \right]}^2}}}{{\ddot R_i^u\left( 0 \right)}}\\
{\rm{   }} = \frac{1}{{0.5\left( {\alpha  + \frac{1}{2}\left( {1 - \alpha } \right){{\left( {1 + \frac{1}{\theta } + \frac{2}{{\sqrt \theta  }}} \right)}^2} + \frac{1}{2}\left( {1 - \alpha } \right){{\left( {1 + \frac{1}{\theta } - \frac{2}{{\sqrt \theta  }}} \right)}^2} - \left( {1 - \alpha } \right)\frac{1}{{{\theta ^2}}}} \right)}} = 1,
\end{array}
\end{equation}
where, $i \in \left\{ {1,2} \right\}$. Hence, the slope region achieved by successive decoding is given by
\begin{equation}
S\left( \theta  \right) = \left\{ {\left( {{{{\cal S}}_1},{{{\cal S}}_2}} \right):0 \le {{{\cal S}}_1} \le 1,0 \le {{{\cal S}}_2} \le 1} \right\}.
\end{equation}

%\clearpage
\bibliographystyle{IEEETCOM}
\bibliography{IEEE_TCOM_V73}

% Generated by IEEEtranTCOM.bst, version: 1.13 (2008/09/30)
\begin{thebibliography}{10}
\baselineskip 12pt
\providecommand{\url}[1]{#1}
\csname url@samestyle\endcsname
\providecommand{\newblock}{\relax}
\providecommand{\bibinfo}[2]{#2}
\providecommand{\BIBentrySTDinterwordspacing}{\spaceskip=0pt\relax}
\providecommand{\BIBentryALTinterwordstretchfactor}{4}
\providecommand{\BIBentryALTinterwordspacing}{\spaceskip=\fontdimen2\font plus
\BIBentryALTinterwordstretchfactor\fontdimen3\font minus
  \fontdimen4\font\relax}
\providecommand{\BIBforeignlanguage}[2]{{%
\expandafter\ifx\csname l@#1\endcsname\relax
\typeout{** WARNING: IEEEtran.bst: No hyphenation pattern has been}%
\typeout{** loaded for the language `#1'. Using the pattern for}%
\typeout{** the default language instead.}%
\else
\language=\csname l@#1\endcsname
\fi
#2}}
\providecommand{\BIBdecl}{\relax}
\BIBdecl

\bibitem{1}
P.~Grover and A.~Sahai, ``Green codes: Energy-efficient short-range
  communication,'' in \emph{IEEE Int. Symp. on Inf. Theory}, July 2008, pp.
  1178--1182.

\bibitem{2}
------, ``Time-division multiplexing for green broadcasting,'' in \emph{IEEE
  Int. Symp. on Inf. Theory}, June 2009, pp. 2517--2521.

\bibitem{3}
K.~Ganesan, P.~Grover, and J.~Rabaey, ``The power cost of over-designing
  codes,'' in \emph{IEEE Workshop on Signal Processing Sys.}, Oct 2011, pp.
  128--133.

\bibitem{4}
J.~Palicot, Y.~Louet, and M.~Mroué, ``Peak to average power ratio sensor for
  green cognitive radio,'' in \emph{IEEE 21st Int. Symp. on Personal Indoor and
  Mobile Radio Commun.}, Sept 2010, pp. 2669--2674.

\bibitem{5}
S.~Verdú, ``On channel capacity per unit cost,'' \emph{IEEE Trans. on Inf.
  Theory}, vol.~36, no.~5, pp. 1019--1030, Sep 1990.

\bibitem{6}
T.~Berger and W.~Levy, ``A mathematical theory of energy efficient neural
  computation and communication,'' \emph{IEEE Trans. on Inf. Theory}, vol.~56,
  no.~2, pp. 852--874, Feb 2010.

\bibitem{7}
P.~Suksompong and T.~Berger, ``Capacity analysis for integrate-and-fire neurons
  with descending action potential thresholds,'' \emph{IEEE Trans. on Inf.
  Theory}, vol.~56, no.~2, pp. 838--851, Feb 2010.

\bibitem{8}
C.~Shannon, ``A mathematical theory of communication,'' \emph{Bell System
  Technical Journal}, vol.~27, no.~4, pp. 623--656, Oct 1948.

\bibitem{9}
R.~Blahut, ``Computation of channel capacity and rate-distortion functions,''
  \emph{IEEE Trans. on Inf. Theory}, vol.~18, no.~4, pp. 460--473, Jul 1972.

\bibitem{10}
S.~Verdu, ``Spectral efficiency in the wideband regime,'' \emph{IEEE Trans. on
  Inf. Theory}, vol.~48, no.~6, pp. 1319--1343, Jun 2002.

\bibitem{11}
A.~Lapidoth, I.~Telatar, and R.~Urbanke, ``On wide-band broadcast channels,''
  \emph{IEEE Trans. on Inf. Theory}, vol.~49, no.~12, pp. 3250--3258, Dec 2003.

\bibitem{12}
G.~Caire, D.~Tuninetti, and S.~Verdu, ``Suboptimality of tdma in the low-power
  regime,'' \emph{IEEE Trans. on Inf. Theory}, vol.~50, no.~4, pp. 608--620,
  April 2004.

\bibitem{13}
A.~El~Gamal, M.~Mohseni, and S.~Zahedi, ``Bounds on capacity and minimum
  energy-per-bit for awgn relay channels,'' \emph{IEEE Trans. on Inf. Theory},
  vol.~52, no.~4, pp. 1545--1561, April 2006.

\bibitem{14}
Y.~Yao, X.~Cai, and G.~Giannakis, ``On energy efficiency and optimum resource
  allocation of relay transmissions in the low-power regime,'' \emph{IEEE
  Trans. on Wireless Commun.}, vol.~4, no.~6, pp. 2917--2927, Nov 2005.

\bibitem{15}
A.~Jain, S.~Kulkarni, and S.~Verdú, ``Minimum energy per bit for gaussian
  broadcast channels with common message and cooperating receivers,'' in
  \emph{47th Annual Allerton Conf. on Commun., Control, and Computing}, Sept
  2009, pp. 740--747.

\bibitem{16}
------, ``Minimum energy per bit for wideband wireless multicasting:
  Performance of decode-and-forward,'' in \emph{Proceedings of IEEE INFOCOM},
  March 2010, pp. 1--9.

\bibitem{17}
A.~Jain, S.~Kulkarni, and S.~Verdu, ``Multicasting in large wireless networks:
  Bounds on the minimum energy per bit,'' \emph{IEEE Trans. on Inf. Theory},
  vol.~57, no.~1, pp. 14--32, Jan 2011.

\bibitem{18}
A.~Host-Madsen, M.~Uppal, and Z.~Xiong, ``On outage capacity in the low power
  regime,'' \emph{IEEE Trans. on Inf. Theory}, vol.~58, no.~2, pp. 888--896,
  Feb 2012.

\bibitem{19}
P.~Banelli, G.~Baruffa, and S.~Cacopardi, ``Effects of hpa nonlinearity on
  frequency multiplexed ofdm signals,'' \emph{IEEE Trans. on Broadcasting},
  vol.~47, no.~2, pp. 123--136, Jun 2001.

\bibitem{20}
S.~Jafar and A.~Goldsmith, ``Adaptive multirate cdma for uplink throughput
  maximization,'' \emph{IEEE Trans. on Wireless Commun.}, vol.~2, no.~2, pp.
  218--228, Mar 2003.

\bibitem{21}
H.~Inaltekin and S.~Hanly, ``Optimality of binary power control for the single
  cell uplink,'' \emph{IEEE Trans. on Inf. Theory}, vol.~58, no.~10, pp.
  6484--6498, Oct 2012.

\bibitem{22}
J.~G. Smith, ``The information capacity of amplitude- and variance-constrained
  sclar gaussian channels,'' \emph{Information and Control}, vol.~18, no.~3,
  pp. 203 -- 219, 1971.

\bibitem{23}
M.~Raginsky, ``On the information capacity of gaussian channels under small
  peak power constraints,'' in \emph{46th Annual Allerton Conf. on Commun.,
  Control, and Computing}, Sept 2008, pp. 286--293.

\bibitem{24}
D.~Guo, S.~Shamai, and S.~Verdu, ``Mutual information and minimum mean-square
  error in gaussian channels,'' \emph{IEEE Trans. on Inf. Theory}, vol.~51,
  no.~4, pp. 1261--1282, April 2005.

\bibitem{25}
V.~Sethuraman and B.~Hajek, ``Low snr capacity of fading channels with peak and
  average power constraints,'' in \emph{IEEE Int. Symp. on Inf. Theory}, July
  2006, pp. 689--693.

\bibitem{26}
A.~Farid and S.~Hranilovic, ``Capacity of optical intensity channels with peak
  and average power constraints,'' in \emph{IEEE Int. Conf. on Commun.}, June
  2009, pp. 1--5.

\bibitem{27}
V.~Sethuraman, L.~Wang, B.~Hajek, and A.~Lapidoth, ``Low-snr capacity of
  noncoherent fading channels,'' \emph{IEEE Trans. on Inf. Theory}, vol.~55,
  no.~4, pp. 1555--1574, April 2009.

\bibitem{28}
T.~Chan, S.~Hranilovic, and F.~Kschischang, ``Capacity-achieving probability
  measure for conditionally gaussian channels with bounded inputs,'' \emph{IEEE
  Trans. on Inf. Theory}, vol.~51, no.~6, pp. 2073--2088, June 2005.

\bibitem{29}
G.~Durisi, U.~Schuster, H.~Bolcskei, and S.~Shamai, ``Noncoherent capacity of
  underspread fading channels,'' \emph{IEEE Trans. on Inf. Theory}, vol.~56,
  no.~1, pp. 367--395, Jan 2010.

\bibitem{30}
S.~Shamai, ``Capacity of a pulse amplitude modulated direct detection photon
  channel,'' \emph{IEE Proceedings I Commun., Speech and Vision}, vol. 137,
  no.~6, pp. 424--430, Dec 1990.

\bibitem{31}
S.~Ikeda and J.~Manton, ``Capacity of a single spiking neuron channel,''
  \emph{Neural Computation}, vol.~21, no.~6, pp. 1714--1748, June 2009.

\bibitem{32}
B.~Mamandipoor, K.~Moshksar, and A.~Khandani, ``On the sum-capacity of gaussian
  mac with peak constraint,'' in \emph{IEEE Int. Symp. on Inf. Theory}, July
  2012, pp. 26--30.

\bibitem{33}
K.~Moshksar, B.~Mamandipoor, and A.~Khandani, ``On orthogonal signalling in
  gaussian multiple access channel with peak constraints,'' in \emph{IEEE Int.
  Symp. on Inf. Theory}, July 2013, pp. 1481--1485.

\bibitem{34}
B.~Mamandipoor, K.~Moshksar, and A.~Khandani, ``Capacity-achieving
  distributions in gaussian multiple access channel with peak power
  constraints,'' \emph{IEEE Trans. on Inf. Theory}, vol.~60, no.~10, pp.
  6080--6092, Oct 2014.

\bibitem{35}
O.~Ozel and S.~Ulukus, ``On the capacity region of the gaussian mac with
  batteryless energy harvesting transmitters,'' in \emph{IEEE Global Commun.
  Conf.}, Dec 2012, pp. 2385--2390.

\bibitem{36}
V.~Sethuraman and B.~Hajek, ``Capacity per unit energy of fading channels with
  a peak constraint,'' \emph{IEEE Trans. on Inf. Theory}, vol.~51, no.~9, pp.
  3102--3120, Sept 2005.

\bibitem{37}
E.~Ordentlich, ``Maximizing the entropy of a sum of independent bounded random
  variables,'' \emph{IEEE Trans. on Inf. Theory}, vol.~52, no.~5, pp.
  2176--2181, May 2006.

\bibitem{38}
R.~W. Yeung, \emph{Information Theory and Network Coding}, 1st~ed.\hskip 1em
  plus 0.5em minus 0.4em\relax Springer Publishing Company, Incorporated, 2008.

\bibitem{40}
S.~Arimoto, ``An algorithm for computing the capacity of arbitrary discrete
  memoryless channels,'' \emph{IEEE Trans. on Inf. Theory}, vol.~18, no.~1, pp.
  14--20, Jan 1972.

\bibitem{41}
M.~Rezaeian and A.~Grant, ``Computation of total capacity for discrete
  memoryless multiple-access channels,'' \emph{IEEE Trans. on Inf. Theory},
  vol.~50, no.~11, pp. 2779--2784, Nov 2004.

\bibitem{42}
A.~Feiten and R.~Mathar, ``Capacity-achieving discrete signaling over additive
  noise channels,'' in \emph{IEEE Int. Conf. on Commun.}, June 2007, pp.
  5401--5405.

\bibitem{43}
T.~M. Cover and J.~A. Thomas, \emph{Elements of Information Theory (Wiley
  Series in Telecommunications and Signal Processing)}.\hskip 1em plus 0.5em
  minus 0.4em\relax Wiley-Interscience, 2006.

\end{thebibliography}

\end{document}